\pdfoutput=1

\documentclass[11pt]{article}

\usepackage[preprint]{acl}

\usepackage{times}
\usepackage{latexsym}
\usepackage{booktabs}
\usepackage{multirow}
\usepackage{graphicx}
\usepackage{xspace}

\usepackage[utf8]{inputenc} 
\usepackage[T1]{fontenc}    
\usepackage{hyperref}       
\usepackage{url}            
\usepackage{amsfonts}       
\usepackage{nicefrac}       
\usepackage{microtype}      
\usepackage{colortbl}
\usepackage{xcolor}         
\usepackage{bm}
\usepackage{amsmath,amssymb}
\usepackage{ulem}
\usepackage{geometry}
\usepackage{adjustbox}
\usepackage{pifont}
\usepackage{tabularx}
\usepackage{booktabs}
\usepackage{makecell}

\usepackage{subcaption}
\usepackage{tikz}
\usepackage{comment}

\usepackage[T1]{fontenc}

\usepackage[utf8]{inputenc}

\usepackage{microtype}

\usepackage{inconsolata}

\title{Codec-SUPERB: An In-Depth Analysis of Sound Codec Models}

\author{Haibin Wu$^{1}$ \thanks{equal first contribution, $^\dag$equal second contribution, order is sorted randomly.}
, Ho-Lam Chung$^{1}\ ^{\star}$, Yi-Cheng Lin$^{1\ \dag}$, Yuan-Kuei Wu$^{1\ \dag}$, Xuanjun Chen$^{1\ \dag}$, \\
\textbf{Yu-Chi Pai$^{1}$, Hsiu-Hsuan Wang$^{1}$, Kai-Wei Chang$^{1}$, Alexander H. Liu$^{2}$, Hung-yi Lee$^{1}$} \\
$^{1}$Graduate Institute of Communication Engineering, National Taiwan University \\
$^{2}$Massachusetts Institute of Technology \\
\texttt{hungyilee@ntu.edu.tw}
}

\begin{document}
\maketitle

\begin{abstract}
The sound codec's dual roles in minimizing data transmission latency and serving as tokenizers underscore its critical importance.
Recent years have witnessed significant developments in codec models.
The ideal sound codec should preserve content, paralinguistics, speakers, and audio information. 
However, the question of which codec achieves optimal sound information preservation remains unanswered, as in different papers, models are evaluated on their selected experimental settings.
This study introduces Codec-SUPERB, an acronym for \textbf{Codec} \textbf{S}ound processing \textbf{U}niversal \textbf{PER}formance \textbf{B}enchmark. 
It is an ecosystem designed to assess codec models across representative sound applications and signal-level metrics rooted in sound domain knowledge.
Codec-SUPERB simplifies result sharing through an online leaderboard, promoting collaboration within a community-driven benchmark database, thereby stimulating new development cycles for codecs.
Furthermore, we undertake an in-depth analysis to offer insights into codec models from both application and signal perspectives, diverging from previous codec papers mainly concentrating on signal-level comparisons.
Finally, we will release codes, the leaderboard, and data to accelerate progress within the community.

\end{abstract}

\begin{figure*}[t]
\centering
\includegraphics[width=6.3in]{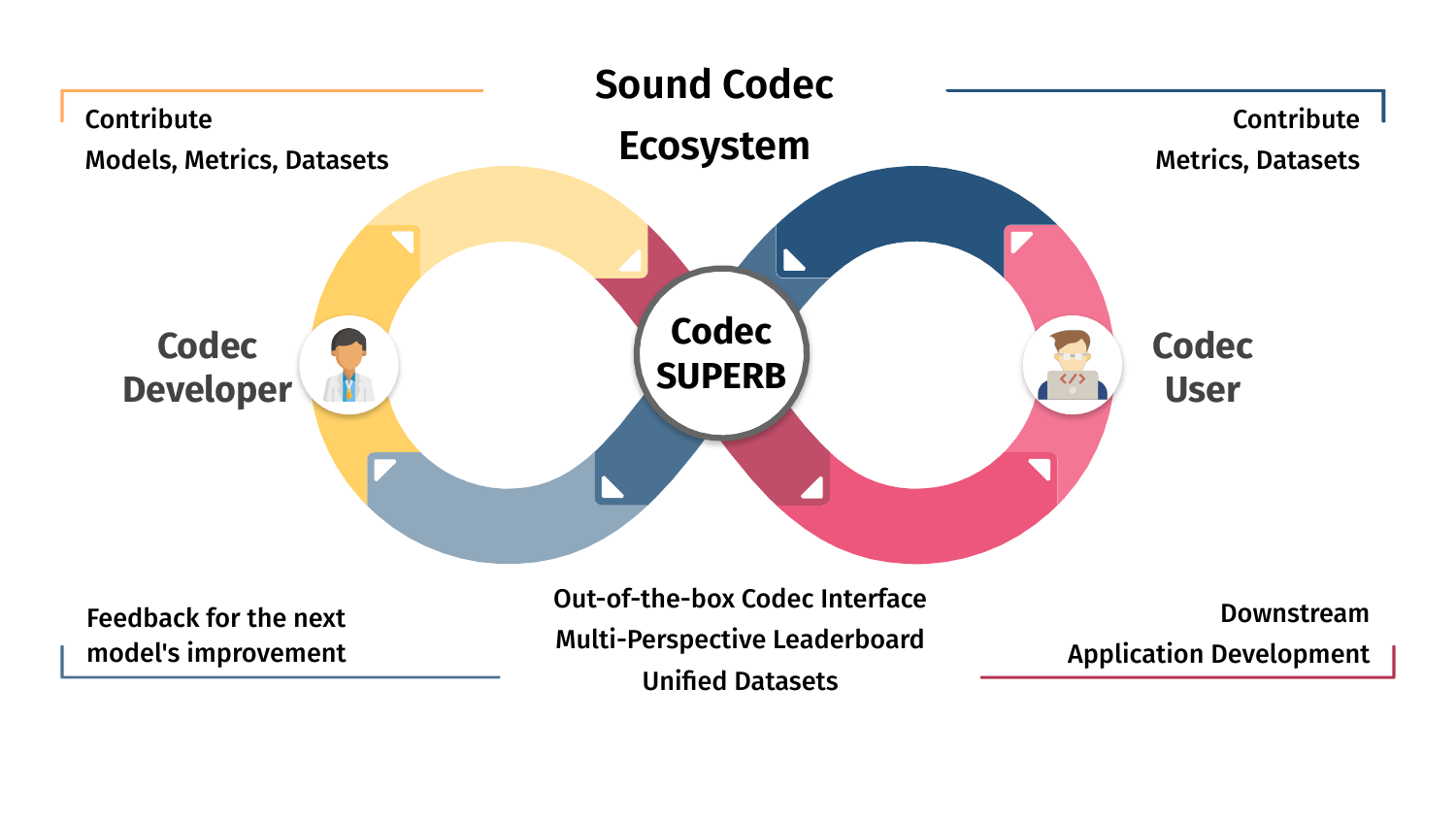}
\caption{
Illustration of the Codec-SUPERB platform from two angles: developers and users.
From the perspective of developers, they develop and evaluate new codec models across a spectrum of sound applications and signal-level metrics defined in our codebase.
Developers then submit their prediction files to the online leaderboard to expand the benchmark database and facilitate comparisons with other codec models.
Ultimately, developers utilize the codebase's visualization and statistical tools to analyze performance discrepancies among Codec-SUPERB applications and metrics, thereby gaining invaluable insights for future improvement directions.
From the users' perspective, they can contribute datasets and metrics and pick codec models for their downstream application usage.
}
\label{fig: platform}
\end{figure*}

\section{Introduction}
\label{sec: intro}

Neural sound codec models were initially introduced to compress sound for efficient data transmission. 
The encoder of the codec model encodes the sound into codec codes, which are then transmitted. Subsequently, the codec decoder then resynthesizes the sound using the received codes.

Neural codec codes can be utilized as tokens in sound language modeling (LM). 
LM has proven highly successful in Natural Language Processing (NLP). 
Sound data contains semantic content and rich information about speaker, emotion, and general audio, offering deeper possibilities for language model applications.
Researchers recently explored the potential of neural codecs 
\citep{defossez2022high, zeghidour2021soundstream, borsos2023soundstorm, wu2023audiodec, yang2023hifi, du2023funcodec, zhang2023speechtokenizer, kumar2023high}
as suitable tokenizers for converting continuous sound into discrete tokens, which can be employed in sound LM \citep{borsos2023audiolm,rubenstein2023audiopalm,agostinelli2023musiclm,wang2023neural,zhang2023speak,wang2023viola,yang2023uniaudio,chen2023lauragpt,wang2023speechx,copet2023simple,lan2023stack,kreuk2022audiogen}. 
Numerous high-performance neural codecs have been developed. 

The dual roles of minimizing data transmission latency and serving as sound LM tokenizers require an ideal codec to preserve content, paralinguistic, speaker, and audio information under low \emph{bitrate} measured by thousand bits per second (\emph{kbps}).
However, the question of which codec achieves the most optimal information preservation across various aspects remains unanswered, as codec models are evaluated with various experimental settings in different papers.
Furthermore, prior codec papers mainly compare performance based on signal-level metrics, neglecting downstream application angles.

To address the aforementioned limitations, we introduce Codec-SUPERB, shorted for \textbf{Codec} \textbf{S}ound processing \textbf{U}niversal \textbf{PER}formance \textbf{B}enchmark, which firstly provides a holistic comparison of the current state-of-the-art codecs under the same, fair and comprehensive experimental setting.
We highlight the following features of Codec-SUPERB:
\begin{enumerate}
    \item \textbf{Diverse angles}: Codec-SUPERB conducts a comprehensive analysis to provide insights into codec models from both application and signal perspectives, diverging from previous codec papers that predominantly focus on signal-level comparisons.
    \item \textbf{Extensive coverage}: Codec-SUPERB exhaustively standardizes the comparison of codec models across six distinct codec models, each with its unique training settings, resulting in 19 distinct codec models. We evaluate 19 codec models across four applications to include comparison for content, speaker, paralinguistic, and audio information. Furthermore, we conduct signal-level comparisons across 20 datasets spanning speech, audio, and music data categories.
    \item \textbf{Community collaboration}: We established an online leaderboard to showcase results, facilitating easy integration of future codec models for public submissions and supporting comparative analysis with statistical and visualization tools (Section~\ref{sec: platform}). We make all resources in Section~\ref{sec: platform} open-source, welcoming researchers to contribute and promote advancements within the codec community.
\end{enumerate}

\section{Codec-SUPERB Platform design}
\label{sec: platform}

As shown in Figure~\ref{fig: platform}, Codec-SUPERB is designed to foster sound codec development by providing a platform for connecting codec developers and codec users. 
Codec-SUPERB is user-friendly for reproducing the model evaluation, assessing the custom codec models, contributing datasets and metrics, and conducting comparative analyses for model characteristics. 
This is facilitated by three core components: an easy-to-follow codebase, a community-driven leaderboard website, and well-selected datasets.

\subsection{Codebase}
\label{subsec: software}
The Codec-SUPERB evaluation processes are conducted through our GitHub repository \footnote{\href{https://anonymous.4open.science/r/Codec-SUPERB-857B/README.md}{Codec-SUPERB codebase}}.
Within this repository, codec models are referred to as \textit{base\_codec} models, intentionally designed to be disentangled from evaluations for downstream applications and signal-level metrics. 
This disentanglement enables users to seamlessly switch between various \textit{base\_codec} and evaluation combinations or add their own \textit{base\_codec} model for evaluation across all sound applications and signal-level metrics. 
The codebase is closely combined with the Codec-SUPERB official leaderboard website \footnote{\href{http://codecsuperb.com}{Codec-SUPERB leaderboard}}, enabling the automatic generation of corresponding submission files upon completion of each codec evaluation.
Users can then upload these submission files, thus effortlessly contributing to expanding the benchmark database.

\begin{figure}[t]
\centering
\includegraphics[width=3.3in]
{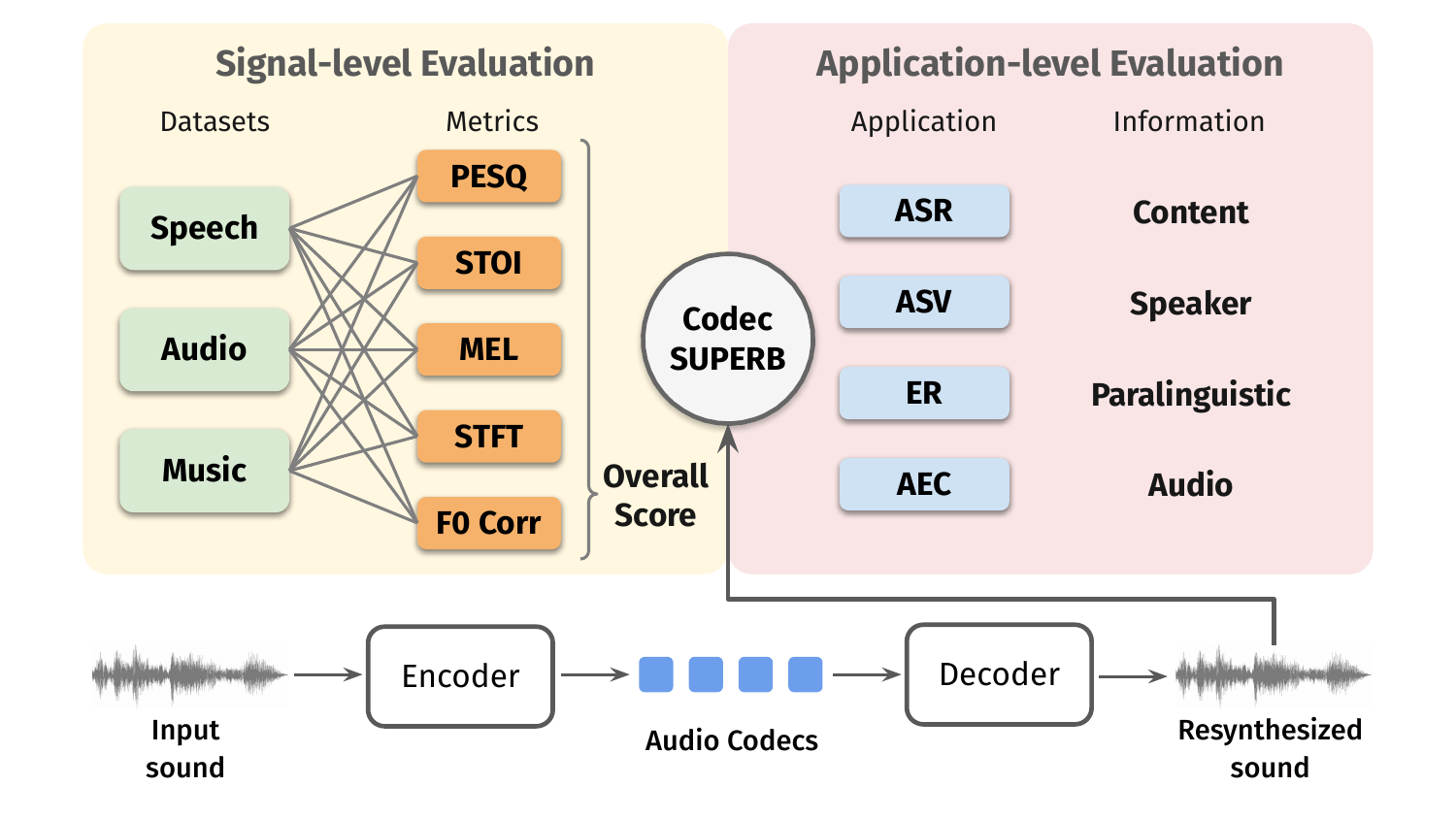}
\vspace{3mm}
\caption{
The input sound is compressed using the codec encoder and resynthesized using the codec decoder. Then the resynthesized sound is evaluated from signal-level and application-level angles. Three categories of dataset, speech, audio, and music, are evaluated using five signal-level metrics and one overall score. Also, 4 downstream applications are evaluated.
}
\label{fig: evaluation process}
\end{figure}

\subsection{Website}
\label{subsec: website}
Our leaderboard website plays a pivotal role in the Codec-SUPERB by continuously expanding the benchmark database, ensuring Codec-SUPERB remains more than just a static leaderboard to showcase our own evaluation results.
Initially, we evaluate 19 codec models and submit them to our online leaderboard.
To lower the participation threshold, the website also accepts submissions with partial results based on the developers' interests from specific angles when evaluating all results is not necessary and cost-prohibitive.
Additionally, the website offers helpful visualization tools for comparing detailed characteristics of different models, as demonstrated in the experiment section.

\subsection{Datasets}
\label{subsec: dataset}
Neural codec models are challenging to compare, even when the source code is available, due to slight variations in evaluation dataset settings. 
This issue is obvious in the sound domain, where differences in sound sampling rates, data partition rules, and waveform preprocessing methods can yield significantly different results. 
To address this challenge and align with sound domain expertise, we curate a comprehensive dataset spanning 20 datasets, comprising a diverse array of speech, music, and audio data. 
These datasets are partitioned according to sound domain knowledge. 
This extensive dataset is publicly accessible and readily available through our leaderboard.

\label{sec:method}

\section{Holistic evaluation in Codec-SUPERB}

Codec models are assessed across diverse experimental settings in various papers. 
Moreover, previous codec studies primarily focus on comparisons using signal-level metrics on their \textit{selected} datasets, overlooking evaluations from downstream application angles.
Thus, we make efforts to address the above limitations by including diverse datasets, comprehensive signal-level metrics, and mainstream sound applications.
The detailed evaluation process is illustrated in Figure~\ref{fig: evaluation process}.

\subsection{Signal-level evaluation}
\label{subsec: Signal-level}

We utilize the codec models to resynthesize the datasets in Section~\ref{subsec:dataset}. Additionally, we employ carefully selected objective metrics outlined in Section~\ref{subsec: objective} and a well-designed overall score detailed in Section~\ref{subsec: overall score} to do signal-level comparisons for different codec models.

\subsubsection{Datasets}
\label{subsec:dataset}
Previous studies typically focus on non-comprehensive categories of data and often rely on a limited number of datasets for evaluation.
We select representative sound datasets spanning three mainstream sound categories: speech, general audio, and music. 
This is because they offer comprehensive perspectives on sound. 
We incorporate all categories across a total of 20 datasets.
To ensure fair comparisons, we standardize the dataset settings, including sampling rate and partition rules. 
All datasets and partition rules are released on our website.
This diverse dataset comprehensively evaluates each codec's performance across various sound types.

\textbf{Speech:} Sound generated by human articulation. It is typically characterized by producing specific sounds and patterns to convey meaningful messages. 
We select speech datasets based on two perspectives: enhancing the diversity of the dataset (i.e., speaker diversity, language variety, and duration) and expanding the range of information preserved in the speech (i.e., emotion and multi-speaker scenarios).

\textbf{Music:} Pattern of sounds created through pitch, tone, and timbre manipulation.
We select music datasets to enhance the diversity of music categories, encompassing various singing voices of different levels of professionalism, music notes played by multiple instruments, and music spanning a variety of genres. 

\textbf{Audio:} Any sound that humans can hear apart from speech and music. 
We chose audio datasets in order to increase the diversity of general audio categories and their applications.

We briefly describe the key information for the selected datasets in Table~\ref{tab:dataset_info}. 
Detailed descriptions, including the partition rules of datasets used in our evaluation, can be found in Appendix~\ref{subsec: Dataset description}.
Also, Table~\ref{tab:dataset_license} in Appendix~\ref{subsec: Dataset description} summarizes the license for each dataset.

\begin{table}[t]
\small
\centering
\resizebox{0.5\textwidth}{!}{%
\begin{tabular}{ll}
\toprule
\textbf{Speech dataset} & \textbf{Features} \\
\midrule
Librispeech  & diverse speaker, read audiobooks \\ 
VoxCeleb1  & diverse speaker, celebrities on YouTube  \\
Speech Commands v1 & spoken keyword commands \\ 
QUESST & multi-lingual, low resource language\\
VoxLingua107 Top 10 & multi-lingual, YouTube content \\
Audio SNIPS & spoken commands, crowdsourced \\
IEMOCAP & affective speech \\
CREMA-D & affective speech \\
Libri2Mix & multi-speaker scenarios \\
LibriCount & multi-speaker scenarios \\
\midrule
\textbf{Audio dataset}& \textbf{Features}  \\
\midrule
ESC-50 & diverse audio source \\
FSD-50K & diverse audio source \\
Gunshot Triangulation & diverse audio source \\
Vocal Imitations & human imitation of sound \\
\midrule
\textbf{Music dataset} & \textbf{Features}  \\
\midrule
OpenSinger & singing voice, Chinese song\\
M4Singer & singing voice, Chinese song \\ 
VocalSet & singing skill \\ 
NSynth & instrument notes \\
GTZAN Genre & diverse music genre \\
GTZAN Music Speech & instrument note \\
\bottomrule
\end{tabular}
}
\caption{Dataset information.}
\label{tab:dataset_info}
\end{table}

\subsubsection{Signal-level metrics}
\label{subsec: objective}

We assess the quality of resynthesized sound using a comprehensive set of signal-level metrics grounded in sound domain expertise. 
These metrics include the Perceptual Evaluation of Speech Quality (PESQ) \citep{rix2001perceptual}, Short-Time Objective Intelligibility (STOI) \citep{taal2010short}, STFT distance (STFTDistance) \citep{alsteris2007short}, Mel distance (MelDistance) \citep{407206}, and F0CORR (F0 Pearson Correlation Coefficient) \cite{parselmouth}. 
The features of the adopted signal-level metrics to assess sound quality are shown in Table~\ref{table:signal_metrics}. 
STFTDistance analyzes frequency content and temporal dynamics, while MelDistance focuses on spectral fidelity and timbral texture, reflecting the Mel scale's relevance to human hearing. 
PESQ provides a subjective quality score, capturing the perceptual quality of speech. 
STOI measures speech intelligibility in noise, essential for clear communication. 
F0CORR evaluates pitch accuracy, crucial for naturalness and expressiveness in sound. 
This diverse set of metrics enables us to conduct a thorough evaluation of sound quality across various dimensions, encompassing spectral fidelity, temporal dynamics, perceptual clarity, and intelligibility.
Details for these metrics are shown in Appendix~\ref{sub_appendix: Objective metrics}

\begin{table}[t]
\centering
\resizebox{0.48\textwidth}{!}{
\begin{tabular}{lp{6cm}p{1.4cm}}
\toprule
\textbf{Metric} & \textbf{Functionality} & \textbf{Range} \\
\midrule
STFTDistance & Frequency content discrepancies. & $[0, \infty)$ \\
\midrule
MelDistance & Gauges the fidelity of spectral features. & $[0, \infty)$ \\
\midrule
PESQ & Rates the perceptual quality of speech. & $[-0.5, 4.5]$ \\
\midrule
STOI & Evaluates speech intelligibility. & $[0, 1]$ \\
\midrule
F0CORR & Measures pitch accuracy. & $[0, 1]$ \\
\bottomrule
\end{tabular}
}
\caption{Summary of signal-level metrics}
\label{table:signal_metrics}
\end{table}


\subsubsection{Overall score for Signal-level metrics}
\label{subsec: overall score}

Currently, no single overall score exists to evaluate signal-level metrics of resynthesized sound produced by codec models. 
What's particularly innovative is our introduction of a unified overall score, which integrates all signal-level metrics in Section~\ref{subsec: objective} for improved visualization. 
Notably, the overall score demonstrates strong correlations with each individual metric as shown in Section~\ref{subsec: Signal-level evaluation}.

The overall score is calculated through normalization and harmonic mean combining all metrics.
Normalization ensures metrics are comparable and less affected by outliers.
For bounded metrics, PESQ, STOI, and F0CORR, we normalize them by subtracting the min and dividing by the range.
For unbounded metrics, STFTDistance and MelDistance, we normalize them by the Sigmoid function.
Inspired by F1 score \citep{chicco2020advantages}, the harmonic mean is used to aggregate the normalized scores, prioritizing balanced performance across metrics.
Similar to the F1 score \citep{chicco2020advantages}, which harmonizes precision and recall, the harmonic mean in our context ensures a balanced evaluation, preventing any single signal-level metric from disproportionately influencing the overall score.



\begin{table*}[t]
\centering
\fontsize{9}{10}\selectfont
\setlength\tabcolsep{3pt}
\begin{tabular}{lcc|ccc|ccccc}
  \toprule
  \multicolumn{3}{c}{(a) Codec Information}& \multicolumn{3}{c}{(b) Signal-level Evaluation} & \multicolumn{5}{c}{(c) Application-level Evaluation} \\
  \midrule
  &\multirow{2}*{\shortstack{\textbf{kbps}}} &  
  \multirow{2}*{\shortstack{\textbf{Other Configuration}}} &  
  \multirow{2}*{\shortstack{\textbf{Speech$\uparrow$}}} &  
  \multirow{2}*{\shortstack{\textbf{Audio$\uparrow$}}} &  
  \multirow{2}*{\shortstack{\textbf{Music$\uparrow$}}} &
  \multirow{2}*{\shortstack{\textbf{WER$\downarrow$}\\ \textbf{(ASR)}}} &
  \multirow{2}*{\shortstack{\textbf{EER$\downarrow$}\\ \textbf{(ASV)}}} &
  \multirow{2}*{\shortstack{\textbf{minDCF$\downarrow$}\\ \textbf{(ASV)}}} &
  \multirow{2}*{\shortstack{\textbf{ACC $\uparrow$}\\ \textbf{(ER)}}} & 
  \multirow{2}*{\shortstack{\textbf{mAP $\uparrow$}\\ \textbf{(AEC)}}}\\
  & & & & & &  &  &  &  &  \\
  \toprule
  None  & -  & -  & -  & - & - & 2.96 & 0.86 & 0.07 & 69.84 & 45.68 \\
  A & 4 & 16k & 0.644 & 0.581 & 0.585 & 4.02 & 3.31 & 0.24 & 65.49 & 15.11 \\
  \midrule
  B1 & 2 & 16k\_320d & 0.610 & 0.574 & 0.601 & 4.94 & 4.43 & 0.29 & 65.96 & 16.19 \\
  B2 & 2 & 16k\_320d\_large\_uni & 0.617 & 0.574 & 0.630  & 6.26 & 5.22 & 0.38 & 64.63 & 28.65 \\
  B3 & 3 & 24k\_320d & 0.611 & 0.592 & 0.604  & 4.49 & 6.16 & 0.36 & 65.95 & 14.01 \\
  \midrule
  C  & 6.4 & 24k\_320d & 0.596 & 0.602 & 0.572 & 3.94 & 5.22 & 0.30 & 65.70 & 17.41 \\
  \midrule
  D1 & 6 & 16k & 0.798 & 0.591 & 0.749 & 3.26 & 1.59 & 0.12 & 68.81 & 41.08 \\
  D2 & 24 & 24k &  \cellcolor{lightgray!20}{\textbf{0.864}} & 0.636 & \cellcolor{lightgray!20}{\textbf{0.815}} & \cellcolor{lightgray!20}{\textbf{2.96}} & 2.24 & 0.14 & \cellcolor{lightgray!20}{\textbf{69.56}} & \cellcolor{lightgray!20}{\textbf{41.37}} \\
  D3 & 8 & 44k & 0.802 & \cellcolor{lightgray!20}{\textbf{0.702}} & 0.770 & 3.18 & 3.59 & 0.26 & 69.18 & 32.04 \\
  \midrule
  E1 & 1.5 & 24k & 0.579 & 0.594 & 0.568 & 9.21 & 13.88 & 0.68 & 58.84 & 18.84 \\
  E2 & 3 & 24k & 0.636 & 0.599 & 0.621 & 4.34 & 6.85 & 0.39 & 63.54 & 26.63 \\
  E3 & 6 & 24k & 0.697 & 0.602 & 0.669 & 3.49 & 4.28 & 0.27 & 66.18 & 32.43 \\
  E4 & 12 & 24k & 0.748 & 0.606 & 0.710 & 3.22 & 3.44 & 0.21 & 67.63 & 35.84 \\
  E5 & 24 & 24k & 0.775 & 0.609 & 0.732 & 3.17 & 3.15 & 0.19 & 68.26 & 36.64 \\
  \midrule
  F1 & 16  & en\_libritts\_16k\_gr1nq32ds320 & 0.724 & 0.582 & 0.667 & 3.21 & \cellcolor{lightgray!20}{\textbf{1.50 }}&\cellcolor{lightgray!20}\textbf{0.10} & 63.54 & 37.31 \\
  F2  & 16 & en\_libritts\_16k\_gr8nq32ds320 & 0.704 & 0.583 & 0.668 & 3.16 & 1.81 & \cellcolor{lightgray!20}{\textbf{0.10}} & 66.18 & 37.77 \\
  F3 & 16 & en\_libritts\_16k\_nq32ds320 & 0.705 & 0.581 & 0.649 & 3.28 & 1.76 & 0.12 & 67.63 & 25.52 \\
  F4 & 8 & en\_libritts\_16k\_nq32ds640 & 0.678 & 0.578 & 0.632 & 3.43 & 2.04 & 0.13 & 68.26 & 21.43 \\
  F5 & 16 & zh\_en\_16k\_nq32ds320 & 0.726 & 0.583 & 0.665 & 3.21 & 1.52 & 0.11 & 69.25 & 26.42 \\
  F6 & 8 & zh\_en\_16k\_nq32ds640& 0.718 & 0.583 & 0.667  & 3.27 & 1.60 & 0.11 & 69.55 & 33.59 \\
  \bottomrule
\end{tabular}%
\caption{Comparison between codec models. (a) Codec information. "A" denotes the Speech Tokenizer, "B$\sim$" signifies the AcademiCodec, "C" is associated with AudioDec, "D$\sim$" represents the DAC, "E$\sim$" refers to the EnCodec, and "F$\sim$" indicates the FunCodec. (b) Signal-level evaluation. (c) Application-level evaluation. "None" means that no codec has been applied.}
\label{tab: codec overall information}
\end{table*}

\subsection{Application-level evaluation}
\label{subsec: downstream}
Beyond previous works mainly focusing on signal-level comparison, we expand our evaluation to include application-level metrics.
This step is essential for comprehensively understanding each codec's ability to preserve crucial sound information, encompassing content, speaker timbre, emotion, and general audio characteristics.
For downstream application evaluation, we utilize pre-trained models to analyze the quality of resynthesized sound. 
Details are shown below.

\subsubsection{Automatic speech recognition (ASR)}
We use ASR to evaluate the content information loss of the codec resynthesis process.
Our study evaluates the ``whisper-large'' variant of the Whisper ASR model \citep{radford2023robust}, renowned for its robust performance across multiple languages and tasks, utilizing an encoder-decoder Transformer architecture. 
We use the most common metric, Word Error Rate (WER), and the most common dataset, LibriSpeech. This evaluation aims to showcase Whisper's proficiency in handling diverse speech qualities and accents, underscoring its potential in real-world speech recognition applications.  
More details can be found in Appendix~\ref{subsub_appendix: ASR}.

\subsubsection{Automatic speaker verification (ASV)}
Speaker information represents a distinct and unique aspect of speech. 
We employ ASV to assess the degree of speaker information loss in the resynthesized speech generated by neural codecs.
As the pre-trained ASV model, we utilize the cutting-edge speaker verification model, ECAPA-TDNN \citep{desplanques2020ecapa}.
We adopt equal error rate (EER) and minimum decision cost function (minDCF) as two evaluation metrics to evaluate the performance of ASV. 
EER provides a balance between false acceptances and rejections, and minDCF allows for a more nuanced assessment of system performance by considering the costs associated with different types of errors (false acceptances and rejections).
More details can be found in Appendix~\ref{subsub_appendix: ASV}

\subsubsection{Emotion recognition (ER)}
In addition to speaker information, speech conveys affective information, including emotions. 
We employ ER to quantify the degree of paralinguistic information loss due to speech resynthesis by codec models.
We utilize the WavLM-Large \cite{chen2022wavlm} self-supervised model for feature extraction and train an emotion classification model on the most famous emotion dataset, IEMOCAP. 
This setting achieves robust and nearly SOTA results.
More details on ER downstream task setting can be found in Appendix~\ref{subsub_appendix: ER} 

\subsubsection{Audio event classification (AEC)}

The goal of adopting AEC is to assess the fidelity of various codecs in preserving audio event information by leveraging a pre-trained AEC model to classify sound events for audio re-synthesized by these codecs. 
We leverage the pre-trained Audio Spectrogram Transformer (AST) \cite{gong21b_interspeech} model and test on the original AudioSet \citep{45857} evaluation set as the baseline. 
More AEC downstream task setting details can be found in Appendix~\ref{subsub_appendix: AEC}. 

\section{Experiments}
\label{sec: experiments}

\subsection{Experimental setup}
\label{subsec: setup}

We adopt six open-source codec models, SpeechTokenizer \citep{zhang2023speechtokenizer}, AudioDec \citep{wu2023audiodec}, AcademiCodec \citep{yang2023hifi}, Descript-audio-codec (DAC) \citep{kumar2023high} Encodec \citep{defossez2022high}, and FunCodec \citep{du2023funcodec}, each with its own distinct training specifications, yielding a total of 19 unique codec models for comparison. 
The column (a) of Table~\ref{tab: codec overall information} provides brief information regarding these models.
Detailed information is in Appendix~\ref{sub_appendix: Codec models}

We select different objective metrics based on the nature of different types of sound to evaluate different categories of sound.
Speech data adopts STFTDistance, MelDistance, PESQ, and STOI.
Audio data adopts STFTDistance and MelDistance.
Music data includes all metrics, particularly F0CORR, for fidelity and expressiveness.

\subsection{Signal-level evaluation}
\label{subsec: Signal-level evaluation}
\begin{figure*}[t]
    \centering
    \begin{minipage}[b]{0.32\textwidth}
        \centering
        \includegraphics[width=\textwidth]{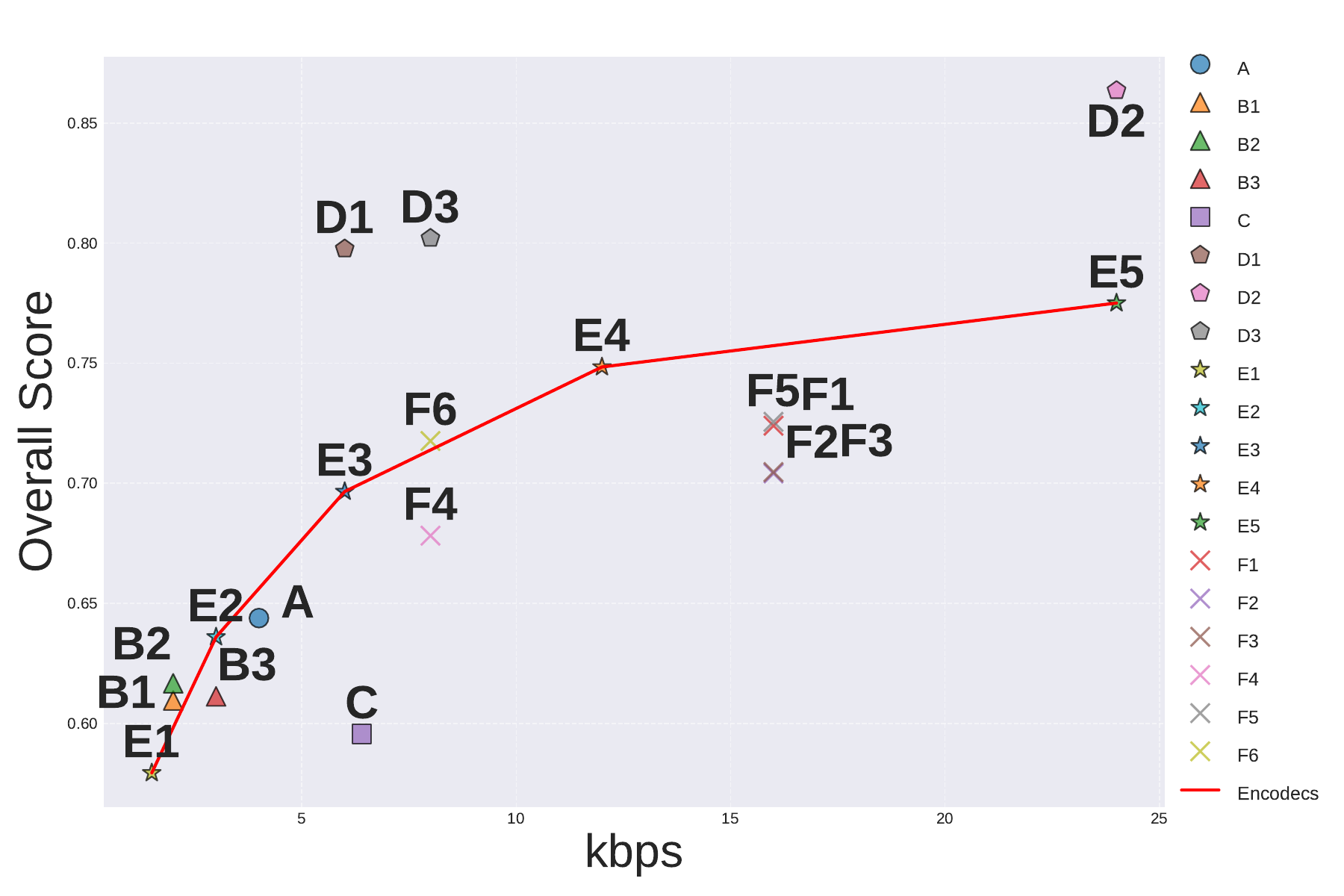}
        \subcaption{Speech Overall Score vs bitrate.}
        \label{fig:suba-overall}
    \end{minipage}
    \hfill
    \begin{minipage}[b]{0.32\textwidth}
        \centering
        \includegraphics[width=\textwidth]{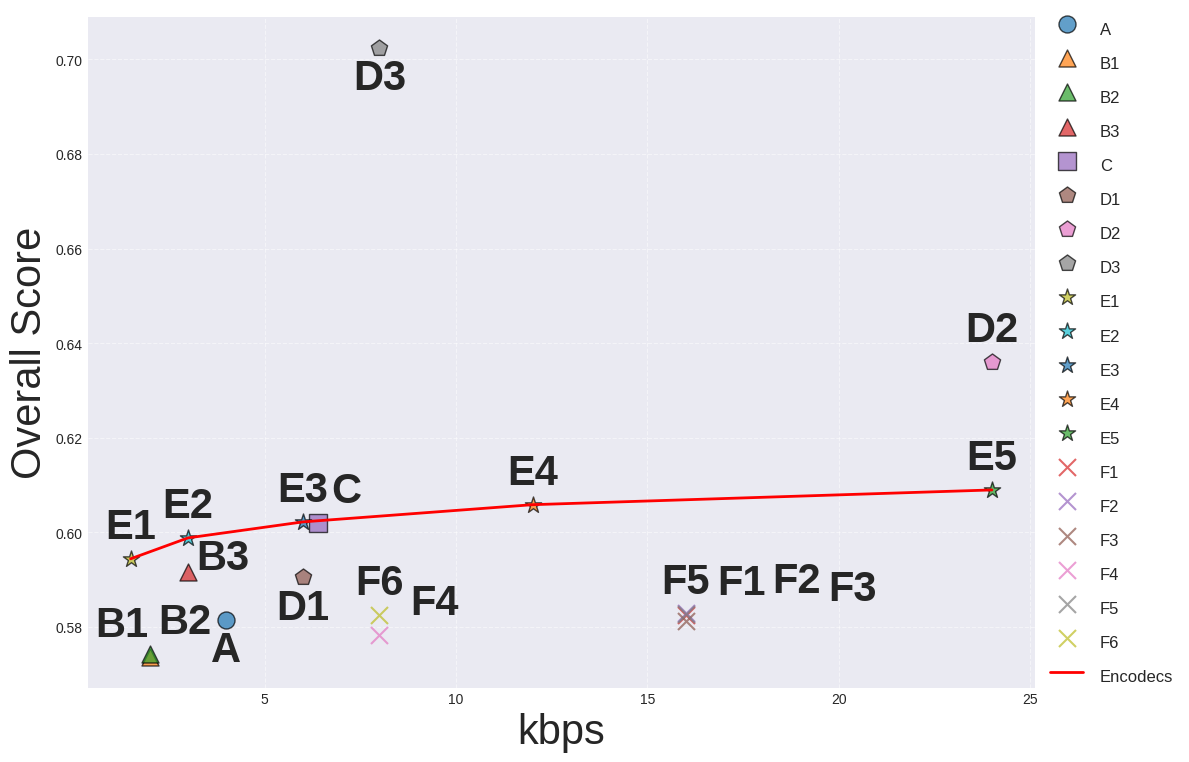}
        \subcaption{Audio Overall Score vs bitrate.}
        \label{fig:subb-overall}
    \end{minipage}
    \hfill
    \begin{minipage}[b]{0.32\textwidth}
        \centering
        \includegraphics[width=\textwidth]{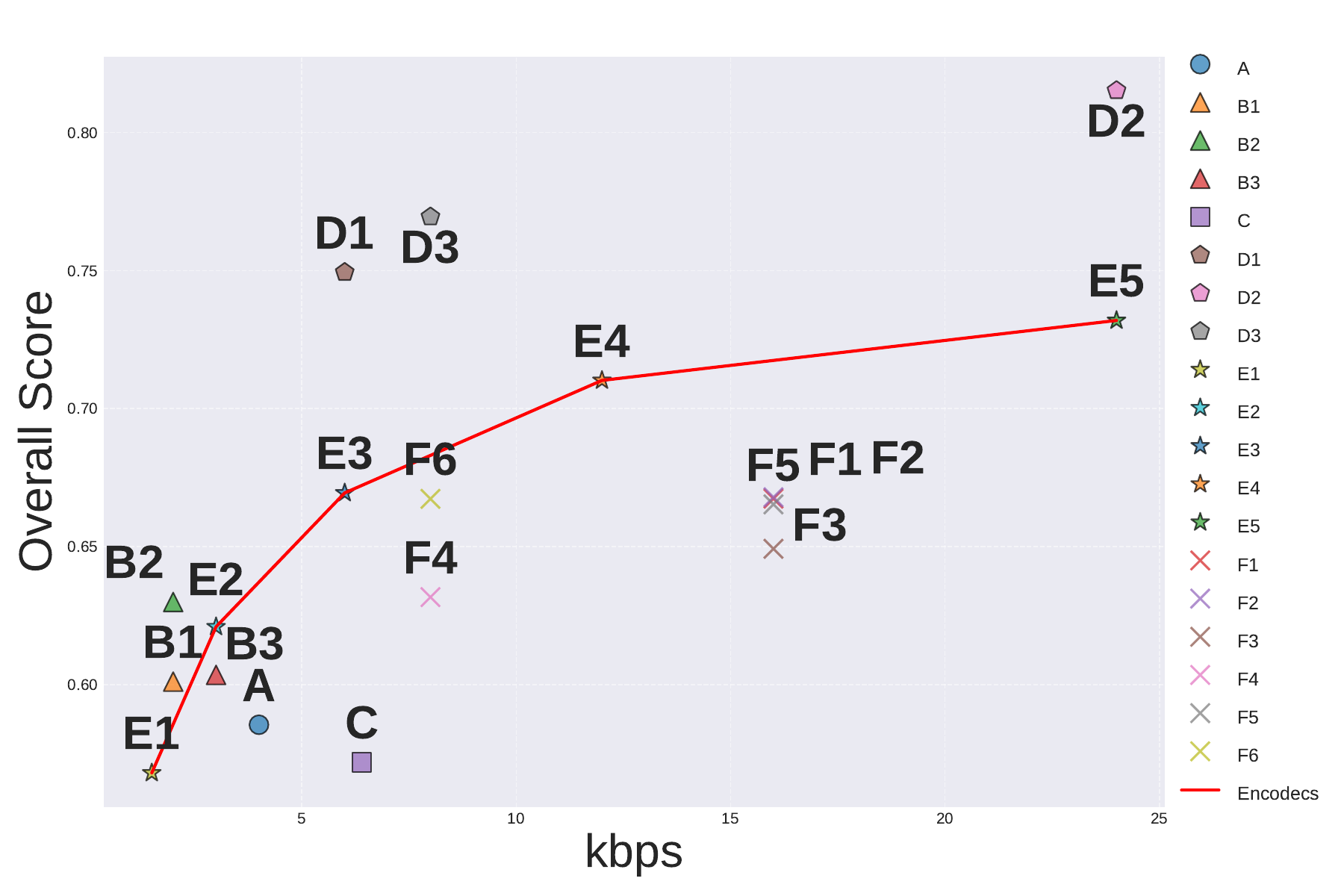}
        \subcaption{Music Overall Score vs bitrate.}
        \label{fig:subc-overall}
    \end{minipage}
    \vspace{3mm}
    \caption{Points in the upper left corner represent a better trade-off between performance and bitrate.}
    \label{fig:fourimages}
\end{figure*}


\begin{table}[t]
\centering
\resizebox{0.5\textwidth}{!}{
\begin{tabular}{lccc}
\toprule
\textbf{Metric} & \textbf{Dataset} & \textbf{Mean Correlation} & \textbf{Mean p-value} \\
\midrule
Mel & Audio & -0.91 & $2.7 \times 10^{-7}$ \\
STFT & Audio & -0.94 & $1.7 \times 10^{-9}$ \\
\midrule
Mel & Music & -0.65 & $2.4 \times 10^{-2}$ \\
STFT & Music & -0.58 & $4.6 \times 10^{-2}$ \\
PESQ & Music & 0.95 & $3.1 \times 10^{-9}$ \\
STOI & Music & 0.83 & $2.8 \times 10^{-3}$ \\
F0CORR & Music & 0.74 & $2.7 \times 10^{-2}$ \\
\midrule
Mel & Speech & -0.77 & $6.1 \times 10^{-4}$ \\
STFT & Speech & -0.71 & $1.0 \times 10^{-2}$ \\
PESQ & Speech & 0.97 & $1.8 \times 10^{-11}$ \\
STOI & Speech & 0.84 & $2.9 \times 10^{-5}$ \\
\bottomrule
\end{tabular}
}
\vspace{3mm}
\caption{Consolidated average correlation coefficients and p-values across three kinds of datasets. A correlation value above 0.7 (below -0.7) indicates a strong positive (negative) correlation. A p-value less than 0.05 denotes significance \cite{corrref}. }
\label{tab:consolidated_correlation_pvalues}
\end{table}


To conduct signal-level evaluation, we employ the ``Overall score'' as the principal metric.
To affirm the overall score as a reliable measure of codec performance to consider diverse signal-level metrics, we conduct a correlation analysis, summarized in Table \ref{tab:consolidated_correlation_pvalues}.
This analysis aims to find the correlation scores between the overall score rankings and those from individual signal-level metric scores. 
We show the mean correlation values for each kind of dataset in Table \ref{tab:consolidated_correlation_pvalues}.
The results of all metrics for all datasets are presented in Figure~\ref{fig:speech overall score correlation} - Figure~\ref{fig:music overall score correlation} in Appendix~\ref{appendix: Additional experiment results}.
Key findings include:
\begin{itemize}
\item MelDistance and STFTDistance have strong negative relations with the overall score.
\item PESQ, STOI, and F0CORR have strong positive relations with the overall score.
\item Mean p-values confirm the above correlations are significant. 
\end{itemize}

This correlation analysis, detailed in Table \ref{tab:consolidated_correlation_pvalues}, establishes the overall score as a comprehensive indicator of codec quality, effectively encompassing various signal-level metrics. 
We can discern which codec achieves superior performance at a given bitrate by comparing the ``Overall Score'' against the bitrate (kbps). 

Table~\ref{tab:combined_overall_scores_speech} to Table~\ref{tab:combined_overall_scores_music} in Appendix~\ref{appendix: Additional experiment results} show the results for each dataset.
We only show the average performance below due to space limitations.
The performance trends are similar

\subsubsection{Speech Dataset}

As shown in Figure~\ref{fig:suba-overall}, for the Speech dataset, it's clear that the Encodec (E1-E5) sets a strong baseline, with only the DAC codec (D1-D3) notably surpassing it at a similar bitrate. 
Other codecs don't show a significant advantage. 
In addition, at very low bitrates, the Academicodec (B1-B3) achieves improved performance.

\subsubsection{Audio Dataset}

As shown in Figure~\ref{fig:subb-overall}, for the Audio dataset, Encodec again proves to be a strong baseline, with DAC being the only codec to surpass its performance significantly. 
Other models do not markedly exceed the performance of Encodec.

\subsubsection{Music Dataset}

As shown in Figure~\ref{fig:subc-overall}, the results within the music dataset seem to consolidate the findings from the previous two datasets. 
Academicodec outperforms Encodec at low bitrates and can even surpass Encodec models when the bitrate is doubled.
DAC also maintains a leading position.

\subsubsection{Takeaways}
The observations across the three datasets indicate that DAC achieves a well-balanced trade-off between performance and bitrate.
In contrast, Academicodec demonstrates the capability to maintain superior performance even at a significantly lower bitrate.
The early-stage model, Encodec, remains a solid baseline.


\begin{figure*}[t]
    \centering
    \begin{subfigure}[b]{0.24\textwidth} 
        \centering
        \includegraphics[width=\textwidth]{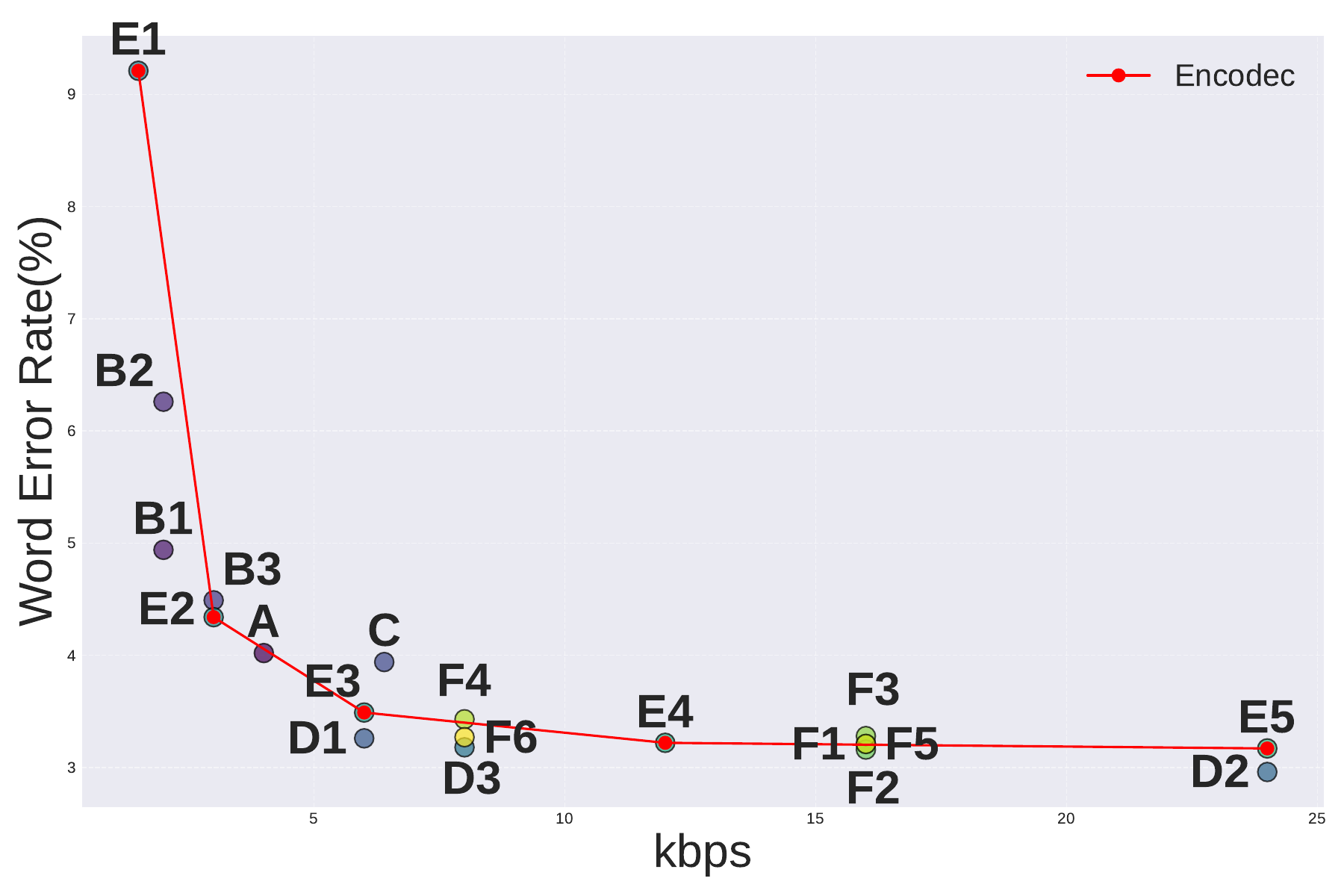}
        \caption{ASR WER (\%) vs bitrate.}
        \label{fig:suba}
    \end{subfigure}
    \hfill
    \begin{subfigure}[b]{0.24\textwidth} 
        \centering
        \includegraphics[width=\textwidth]{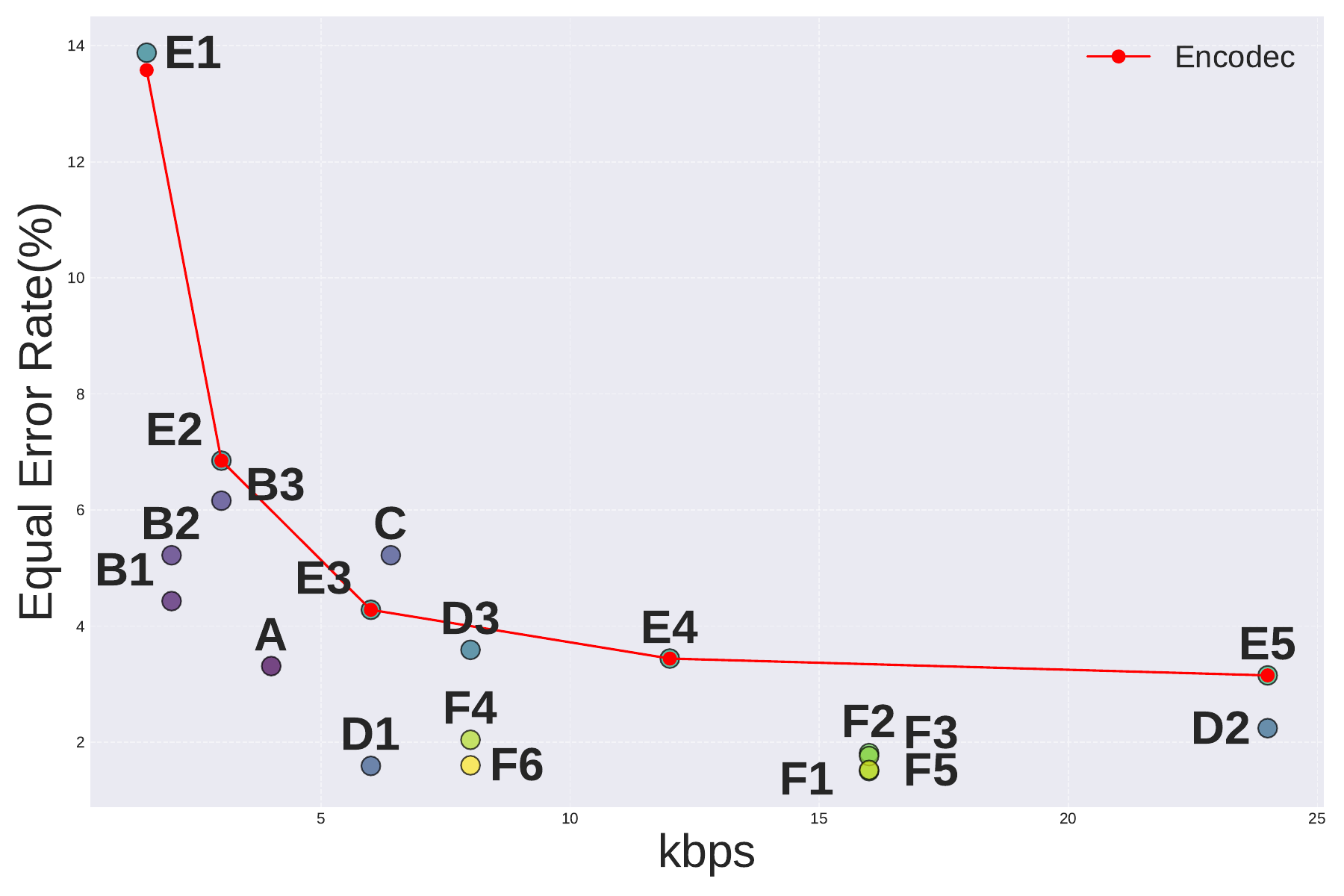}
        \caption{ASV EER (\%) vs bitrate.}
        \label{fig:subb}
    \end{subfigure}
    \hfill
    \begin{subfigure}[b]{0.24\textwidth} 
        \centering
        \includegraphics[width=\textwidth]{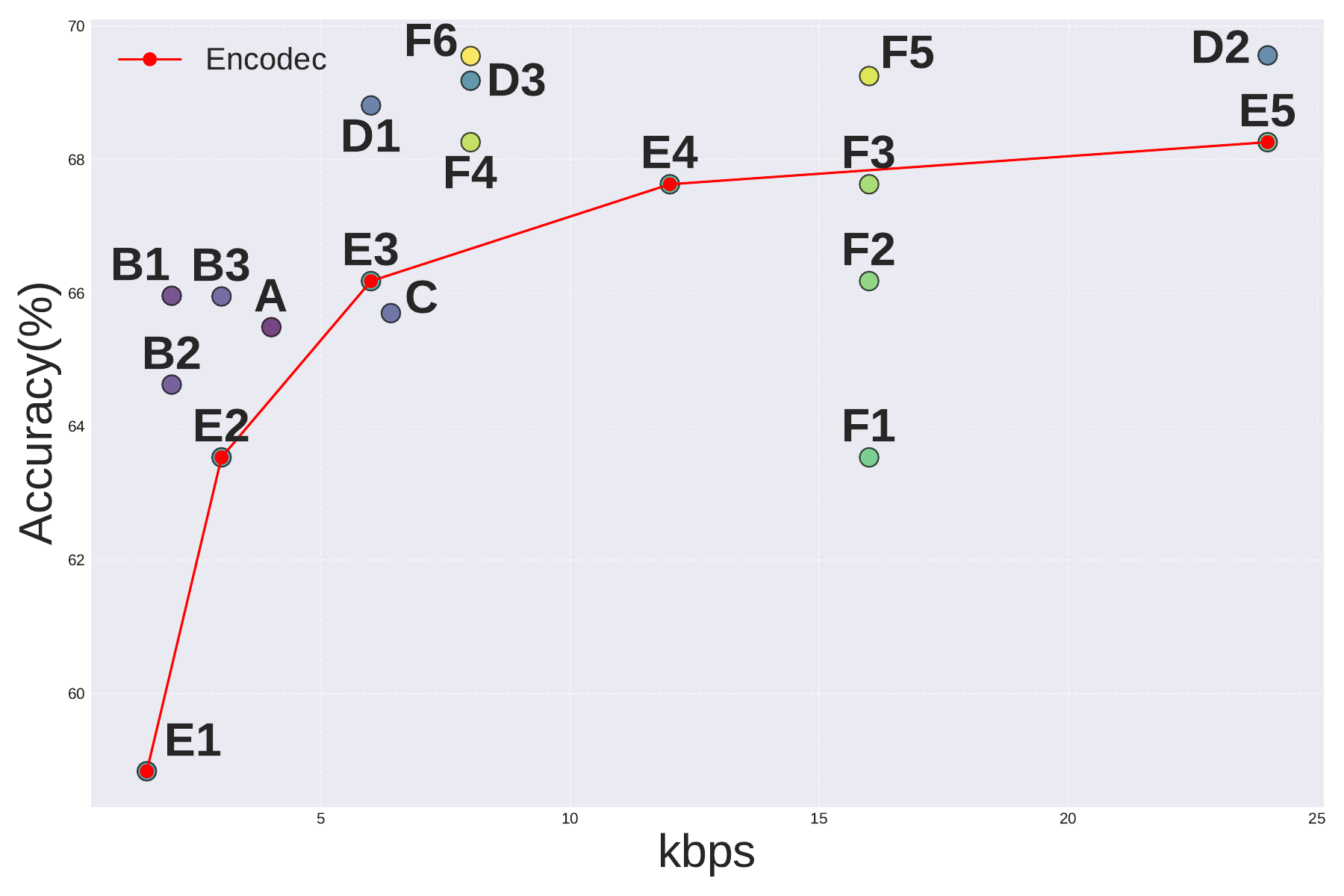}
        \caption{ER ACC (\%) vs bitrate.}
        \label{fig:subc}
    \end{subfigure}
    \hfill
    \begin{subfigure}[b]{0.24\textwidth} 
        \centering
        \includegraphics[width=\textwidth]{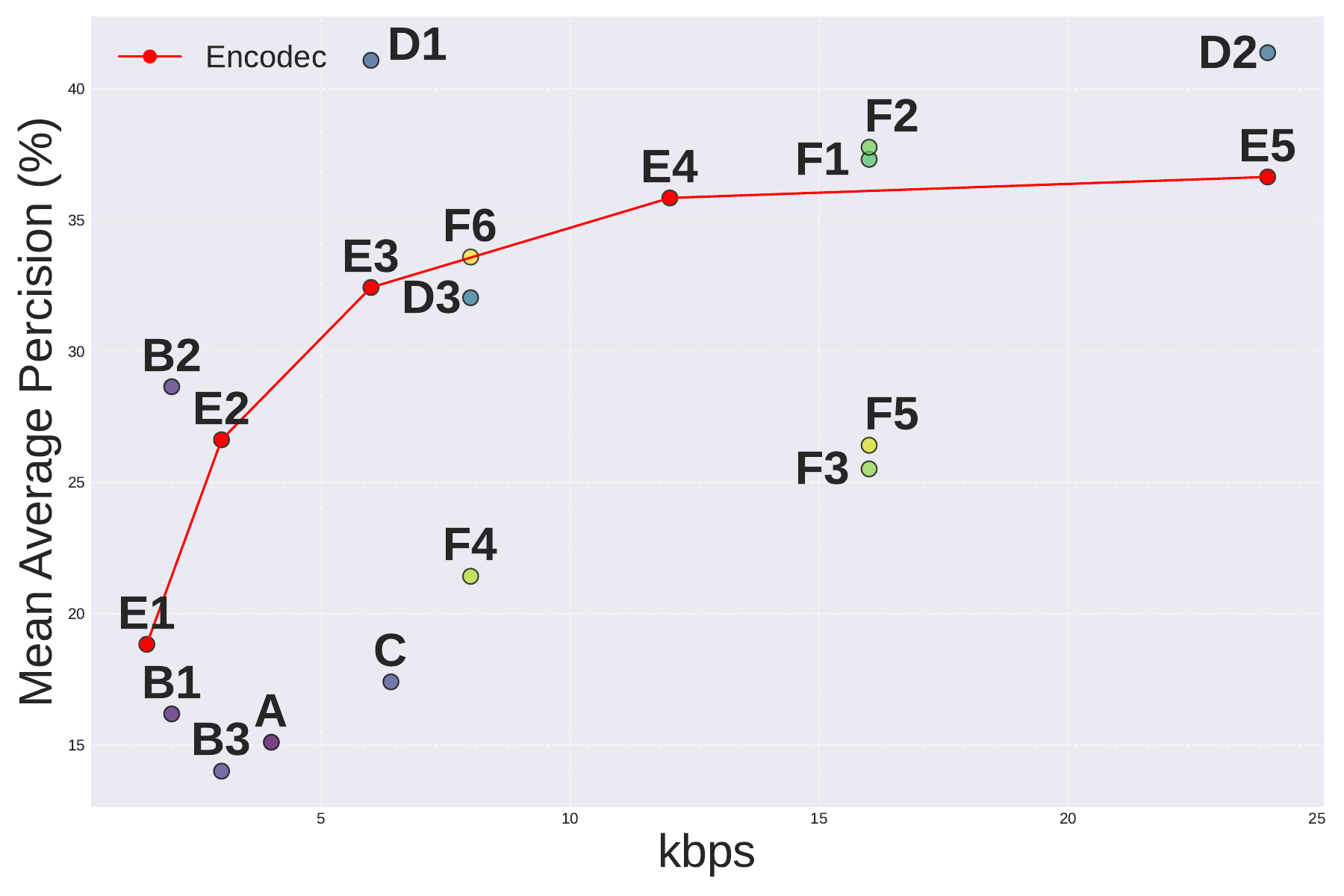}
        \caption{AEC mAP (\%) vs bitrate.}
        \label{fig:subd}
    \end{subfigure}
    \vspace{3mm}
    \caption{
    The trade-off between bitrate and performance of different applications. 
    Models located in the lower/upper left corner of (a)(b)/(c)(d) indicate a more favorable trade-off between performance and bitrates.
    }
    \label{fig:exp-overall-score}
\end{figure*}

\subsection{Application-level evaluation}
\label{subsec: Application-level evaluation}

\subsubsection{Automatic speech recognition}
The column (c) of Table~\ref{tab: codec overall information} (Table~\ref{tab: codec overall information}-c) indicates that the process of codec resynthesis process typically leads to a loss of contextual information in speech, adversely affecting the Word Error Rate (WER). 
However, an intriguing exception to this pattern is observed with the D2 codec (dac\_24k), which maintains ASR performance comparable to the original, unprocessed speech. This suggests that the D2 codec's resynthesis process uniquely preserves the integrity of the speech content.
As depicted in Figure~\ref{fig:suba}, we observe that:
(1) WER consistently decreases as bitrate increases, signifying that a higher bitrate contributes to preserving content information within the codec;
(3) the DAC codecs(D1-D3) exhibit the lowest WER across the board, indicating their effectiveness in maintaining content information during resynthesis;
(4) in contrast, with a bitrate of around 6kbps, the AudioDec codec (C) obtained the highest WER, which indicates a significant loss of content information.

\subsubsection{Automatic speaker verification}
As presented in Table~\ref{tab: codec overall information}-c, adopting codecs A to F6 to the original audio leads to some loss of speaker information. 
Based on Figure~\ref{fig:subb} and Table~~\ref{tab: codec overall information}-c, we can observe that:
(1) when comparing Encodec E1 to E5, we observe an increase in bitrate and a decrease in EER, indicating that a higher bitrate can better preserve speaker information;
(2) Funcodec F2 has the lowest EER and minDCF, resulting in the least degradation of speaker information.
Funcodec F2 also attains the highest bitrate, 
which results in preserving more information than other codec models;
(3) B1, A, and D1 attained the optimal Pareto balance, effectively striking a fine trade-off between EER and bitrate;
(4) DAC D1 attained an impressively low EER while maintaining a reasonable bitrate.

\subsubsection{Emotion recognition}
Our observations are drawn from Figure~\ref{fig:subc} and Table~\ref{tab: codec overall information}-c:
(1) when comparing Encodec from E1 to E5, we observe an increase in bitrate and Accuracy (\%), indicating that a higher bitrate can preserve more information;
(2) DAC model D2 has the highest ACC 69.56\%. D2 only drops in accuracy by 0.38\% compared to the original audio;
(3) Encodec E1 has the worst accuracy. When comparing E1 with B1 and B2, both having similar bitrate, a significant decrease in accuracy is observed;
(4) under the same bitrate, DAC outperforms all the other codecs; F2 outperforms F1, F3, and F5; AcademiCodec B1 outperforms B2 even though they have the same architecture while B2 is trained on a larger set of data; AcademiCodec B3 outperforms Encodec E3;
(5) in general, Encodec models and AudioDec have the lowest accuracy under a similar bitrate, suggesting more loss of emotional information;
(6) even for the least performing model, E1, with a bitrate of only 1.5kbps, it can still maintain the accuracy drop from 69.84\% to 58.84\%. 
This implies that emotional information can be retained even with a very low bitrate.

\subsubsection{Audio event classification}
\label{subsubsec_Analysis_AEC}

Based on Figure~\ref{fig:subd} and Table~\ref{tab: codec overall information}-c, we can observe that: 
(1) when comparing Encodec from E1 to E5, we observe an increase in bitrate and mAP (\%), indicating that a higher bitrate can better preserve audio event information; 
(2) DAC model D2 has the highest mAP 41.37\%, resulting in the least degradation of audio event information. 
(3) EnCodec and DAC models are trained using AudioSet training set, so when comparing mAP (\%) for AudioSet testing set under different bitrates, EnCodec often exceeds SpeechTokenizer (around 4kbps), AcademiCodec (around 3kbps), and AudioDec (around 6kbps);
(4) AcademiCodec B2 significantly surpassed EnCodec at a similar rate as it is trained on a diverse speech dataset.
(5) FunCodec was not trained using the AudioSet training set. 
When comparing model F6 with F4, it is evident that F6's performance closely approaches that of EnCodec (which is trained using Audioset). 
The primary distinction between them is that F6 utilizes a multi-domain speech dataset, whereas F4 relies solely on LibriTTS \citep{zen2019libritts}. 

\subsubsection{Takeaways}
Here we summarize our findings:
(1) emotion information can be conserved even at a remarkably low bitrate of 1.5kbps;
(2) despite not being trained on Audioset, some models (F6) trained on diverse enough speech data can generalize and effectively maintain audio information;
(3) there exists a clear trade-off between bitrate and the quality of codec resynthesis in terms of all downstream tasks we covered;
(4) among the higher bitrate (6kbps$\sim$24kbps) models, DAC outperforms other codecs in retaining content, emotion, speaker, and audio information under similar bitrate;
(5) the best low-bitrate model is Academicodec, which performs excellently in preserving content, emotion, speaker, and audio information from 2 - 3kpbs.

\section{Limitations}

Our evaluation data spans 20 datasets, each containing numerous testing samples, resulting in a substantial amount of data that requires significant computation time and resources. 
As a solution, we aim to devise criteria for identifying representative data points within each dataset to expedite the evaluation process.

\section{Conclusion}
This study presents Codec-SUPERB, a public framework tailored to fairly and comprehensively evaluate codec models. 
Comprised of a user-friendly codebase, a collaborative benchmark driven by the community, and meticulously crafted metrics paired with curated datasets, Codec-SUPERB streamlines result comparisons through an interactive online leaderboard. 
Additionally, our comprehensive analysis provides valuable insights into codec models, examining both application and signal perspectives. 
This departure from prior codec research, primarily focusing on signal-level comparisons, allows for a richer understanding of codec performance. 
Our innovative introduction of a unified overall score sets us apart from previous works, seamlessly integrating all signal-level metrics to enhance visualization. 
Remarkably, this overall score exhibits robust correlations with each individual metric, underscoring its reliability.
Lastly, we commit to releasing codes, the leaderboard, and data resources to expedite progress and foster growth within the community.

\section{Acknowledgment}
We thank the National Center for High-performance Computing (NCHC) of the National Applied Research Laboratories (NARLabs) in Taiwan for providing computational and storage resources.

\newpage
\bibliography{main}

\begin{thebibliography}{64}
\expandafter\ifx\csname natexlab\endcsname\relax\def\natexlab#1{#1}\fi

\bibitem[{cor(2015)}]{corrref}
 2015.
\newblock \href {https://doi.org/10.4135/9781473948167} {Learn about pearson's correlation coefficient in spss with data from the global health observatory data (2012)}.

\bibitem[{Agostinelli et~al.(2023)Agostinelli, Denk, Borsos, Engel, Verzetti, Caillon, Huang, Jansen, Roberts, Tagliasacchi et~al.}]{agostinelli2023musiclm}
Andrea Agostinelli, Timo~I Denk, Zal{\'a}n Borsos, Jesse Engel, Mauro Verzetti, Antoine Caillon, Qingqing Huang, Aren Jansen, Adam Roberts, Marco Tagliasacchi, et~al. 2023.
\newblock Musiclm: Generating music from text.
\newblock \emph{arXiv preprint arXiv:2301.11325}.

\bibitem[{Alsteris and Paliwal(2007)}]{alsteris2007short}
Leigh~D Alsteris and Kuldip~K Paliwal. 2007.
\newblock Short-time phase spectrum in speech processing: A review and some experimental results.
\newblock \emph{Digital signal processing}, 17(3):578--616.

\bibitem[{Anguera et~al.(2015)Anguera, Rodriguez-Fuentes, Buzo, Metze, Sz{\"o}ke, and Penagarikano}]{anguera2015quesst2014}
Xavier Anguera, Luis-J Rodriguez-Fuentes, Andi Buzo, Florian Metze, Igor Sz{\"o}ke, and Mikel Penagarikano. 2015.
\newblock Quesst2014: Evaluating query-by-example speech search in a zero-resource setting with real-life queries.
\newblock In \emph{2015 IEEE International Conference on Acoustics, Speech and Signal Processing (ICASSP)}, pages 5833--5837. IEEE.

\bibitem[{Borsos et~al.(2023{\natexlab{a}})Borsos, Marinier, Vincent, Kharitonov, Pietquin, Sharifi, Roblek, Teboul, Grangier, Tagliasacchi et~al.}]{borsos2023audiolm}
Zal{\'a}n Borsos, Rapha{\"e}l Marinier, Damien Vincent, Eugene Kharitonov, Olivier Pietquin, Matt Sharifi, Dominik Roblek, Olivier Teboul, David Grangier, Marco Tagliasacchi, et~al. 2023{\natexlab{a}}.
\newblock Audiolm: a language modeling approach to audio generation.
\newblock \emph{IEEE/ACM Transactions on Audio, Speech, and Language Processing}.

\bibitem[{Borsos et~al.(2023{\natexlab{b}})Borsos, Sharifi, Vincent, Kharitonov, Zeghidour, and Tagliasacchi}]{borsos2023soundstorm}
Zal{\'a}n Borsos, Matt Sharifi, Damien Vincent, Eugene Kharitonov, Neil Zeghidour, and Marco Tagliasacchi. 2023{\natexlab{b}}.
\newblock Soundstorm: Efficient parallel audio generation.
\newblock \emph{arXiv preprint arXiv:2305.09636}.

\bibitem[{Busso et~al.(2008)Busso, Bulut, Lee, Kazemzadeh, Mower, Kim, Chang, Lee, and Narayanan}]{busso2008iemocap}
Carlos Busso, Murtaza Bulut, Chi-Chun Lee, Abe Kazemzadeh, Emily Mower, Samuel Kim, Jeannette~N Chang, Sungbok Lee, and Shrikanth~S Narayanan. 2008.
\newblock Iemocap: Interactive emotional dyadic motion capture database.
\newblock \emph{Language resources and evaluation}, 42:335--359.

\bibitem[{Cao et~al.(2014)Cao, Cooper, Keutmann, Gur, Nenkova, and Verma}]{6849440}
Houwei Cao, David~G. Cooper, Michael~K. Keutmann, Ruben~C. Gur, Ani Nenkova, and Ragini Verma. 2014.
\newblock \href {https://doi.org/10.1109/TAFFC.2014.2336244} {Crema-d: Crowd-sourced emotional multimodal actors dataset}.
\newblock \emph{IEEE Transactions on Affective Computing}, 5(4):377--390.

\bibitem[{Chen et~al.(2023)Chen, Chu, Gao, Li, Hu, Zhou, Xu, Ma, Wang, Zheng et~al.}]{chen2023lauragpt}
Qian Chen, Yunfei Chu, Zhifu Gao, Zerui Li, Kai Hu, Xiaohuan Zhou, Jin Xu, Ziyang Ma, Wen Wang, Siqi Zheng, et~al. 2023.
\newblock Lauragpt: Listen, attend, understand, and regenerate audio with gpt.
\newblock \emph{arXiv preprint arXiv:2310.04673}.

\bibitem[{Chen et~al.(2022)Chen, Wang, Chen, Wu, Liu, Chen, Li, Kanda, Yoshioka, Xiao et~al.}]{chen2022wavlm}
Sanyuan Chen, Chengyi Wang, Zhengyang Chen, Yu~Wu, Shujie Liu, Zhuo Chen, Jinyu Li, Naoyuki Kanda, Takuya Yoshioka, Xiong Xiao, et~al. 2022.
\newblock Wavlm: Large-scale self-supervised pre-training for full stack speech processing.
\newblock \emph{IEEE Journal of Selected Topics in Signal Processing}, 16(6):1505--1518.

\bibitem[{Chicco and Jurman(2020)}]{chicco2020advantages}
Davide Chicco and Giuseppe Jurman. 2020.
\newblock The advantages of the matthews correlation coefficient (mcc) over f1 score and accuracy in binary classification evaluation.
\newblock \emph{BMC genomics}, 21(1):1--13.

\bibitem[{Chung et~al.(2018)Chung, Nagrani, and Zisserman}]{chung2018voxceleb2}
Joon~Son Chung, Arsha Nagrani, and Andrew Zisserman. 2018.
\newblock Voxceleb2: Deep speaker recognition.
\newblock \emph{arXiv preprint arXiv:1806.05622}.

\bibitem[{Cooper and Shaw(2020)}]{cooper2020gunshots}
Seth Cooper and Steven Shaw. 2020.
\newblock Gunshots recorded in an open field using ipod touch devices.
\newblock \emph{Dryad, Dataset}.

\bibitem[{Copet et~al.(2023)Copet, Kreuk, Gat, Remez, Kant, Synnaeve, Adi, and D{\'e}fossez}]{copet2023simple}
Jade Copet, Felix Kreuk, Itai Gat, Tal Remez, David Kant, Gabriel Synnaeve, Yossi Adi, and Alexandre D{\'e}fossez. 2023.
\newblock Simple and controllable music generation.
\newblock \emph{arXiv preprint arXiv:2306.05284}.

\bibitem[{Cosentino et~al.(2020)Cosentino, Pariente, Cornell, Deleforge, and Vincent}]{cosentino2020librimix}
Joris Cosentino, Manuel Pariente, Samuele Cornell, Antoine Deleforge, and Emmanuel Vincent. 2020.
\newblock Librimix: An open-source dataset for generalizable speech separation.
\newblock \emph{arXiv preprint arXiv:2005.11262}.

\bibitem[{D{\'e}fossez et~al.(2022)D{\'e}fossez, Copet, Synnaeve, and Adi}]{defossez2022high}
Alexandre D{\'e}fossez, Jade Copet, Gabriel Synnaeve, and Yossi Adi. 2022.
\newblock High fidelity neural audio compression.
\newblock \emph{arXiv preprint arXiv:2210.13438}.

\bibitem[{Desplanques et~al.(2020)Desplanques, Thienpondt, and Demuynck}]{desplanques2020ecapa}
Brecht Desplanques, Jenthe Thienpondt, and Kris Demuynck. 2020.
\newblock Ecapa-tdnn: Emphasized channel attention, propagation and aggregation in tdnn based speaker verification.
\newblock \emph{arXiv preprint arXiv:2005.07143}.

\bibitem[{Du et~al.(2023)Du, Zhang, Hu, and Zheng}]{du2023funcodec}
Zhihao Du, Shiliang Zhang, Kai Hu, and Siqi Zheng. 2023.
\newblock Funcodec: A fundamental, reproducible and integrable open-source toolkit for neural speech codec.
\newblock \emph{arXiv preprint arXiv:2309.07405}.

\bibitem[{Engel et~al.(2017)Engel, Resnick, Roberts, Dieleman, Norouzi, Eck, and Simonyan}]{pmlr-v70-engel17a}
Jesse Engel, Cinjon Resnick, Adam Roberts, Sander Dieleman, Mohammad Norouzi, Douglas Eck, and Karen Simonyan. 2017.
\newblock \href {https://proceedings.mlr.press/v70/engel17a.html} {Neural audio synthesis of musical notes with {W}ave{N}et autoencoders}.
\newblock In \emph{Proceedings of the 34th International Conference on Machine Learning}, volume~70 of \emph{Proceedings of Machine Learning Research}, pages 1068--1077. PMLR.

\bibitem[{Fonseca et~al.(2022)Fonseca, Favory, Pons, Font, and Serra}]{fonseca2022FSD50K}
Eduardo Fonseca, Xavier Favory, Jordi Pons, Frederic Font, and Xavier Serra. 2022.
\newblock {FSD50K}: an open dataset of human-labeled sound events.
\newblock \emph{IEEE/ACM Transactions on Audio, Speech, and Language Processing}, 30:829--852.

\bibitem[{Gemmeke et~al.(2017)Gemmeke, Ellis, Freedman, Jansen, Lawrence, Moore, Plakal, and Ritter}]{45857}
Jort~F. Gemmeke, Daniel P.~W. Ellis, Dylan Freedman, Aren Jansen, Wade Lawrence, R.~Channing Moore, Manoj Plakal, and Marvin Ritter. 2017.
\newblock Audio set: An ontology and human-labeled dataset for audio events.
\newblock In \emph{Proc. IEEE ICASSP 2017}, New Orleans, LA.

\bibitem[{Gong et~al.(2021)Gong, Chung, and Glass}]{gong21b_interspeech}
Yuan Gong, Yu-An Chung, and James Glass. 2021.
\newblock \href {https://doi.org/10.21437/Interspeech.2021-698} {{AST: Audio Spectrogram Transformer}}.
\newblock In \emph{Proc. Interspeech 2021}, pages 571--575.

\bibitem[{Hochreiter and Schmidhuber(1997)}]{hochreiter1997long}
Sepp Hochreiter and J{\"u}rgen Schmidhuber. 1997.
\newblock Long short-term memory.
\newblock \emph{Neural computation}, 9(8):1735--1780.

\bibitem[{Hsu et~al.(2021)Hsu, Bolte, Tsai, Lakhotia, Salakhutdinov, and Mohamed}]{hsu2021hubert}
Wei-Ning Hsu, Benjamin Bolte, Yao-Hung~Hubert Tsai, Kushal Lakhotia, Ruslan Salakhutdinov, and Abdelrahman Mohamed. 2021.
\newblock Hubert: Self-supervised speech representation learning by masked prediction of hidden units.
\newblock \emph{IEEE/ACM Transactions on Audio, Speech, and Language Processing}, 29:3451--3460.

\bibitem[{Huang et~al.(2021)Huang, Chen, Ren, Liu, Cui, and Zhao}]{huang2021multi}
Rongjie Huang, Feiyang Chen, Yi~Ren, Jinglin Liu, Chenye Cui, and Zhou Zhao. 2021.
\newblock Multi-singer: Fast multi-singer singing voice vocoder with a large-scale corpus.
\newblock In \emph{Proceedings of the 29th ACM International Conference on Multimedia}, pages 3945--3954.

\bibitem[{Jadoul et~al.(2018)Jadoul, Thompson, and de~Boer}]{parselmouth}
Yannick Jadoul, Bill Thompson, and Bart de~Boer. 2018.
\newblock \href {https://doi.org/https://doi.org/10.1016/j.wocn.2018.07.001} {Introducing {P}arselmouth: A {P}ython interface to {P}raat}.
\newblock \emph{Journal of Phonetics}, 71:1--15.

\bibitem[{Kim et~al.(2018)Kim, Ghei, Pardo, and Duan}]{kim2018vocal}
Bongjun Kim, Madhav Ghei, Bryan Pardo, and Zhiyao Duan. 2018.
\newblock Vocal imitation set: a dataset of vocally imitated sound events using the audioset ontology.
\newblock In \emph{DCASE}, pages 148--152.

\bibitem[{Kong et~al.(2020)Kong, Kim, and Bae}]{kong2020hifi}
Jungil Kong, Jaehyeon Kim, and Jaekyoung Bae. 2020.
\newblock Hifi-gan: Generative adversarial networks for efficient and high fidelity speech synthesis.
\newblock \emph{Advances in Neural Information Processing Systems}, 33:17022--17033.

\bibitem[{Kreuk et~al.(2022)Kreuk, Synnaeve, Polyak, Singer, D{\'e}fossez, Copet, Parikh, Taigman, and Adi}]{kreuk2022audiogen}
Felix Kreuk, Gabriel Synnaeve, Adam Polyak, Uriel Singer, Alexandre D{\'e}fossez, Jade Copet, Devi Parikh, Yaniv Taigman, and Yossi Adi. 2022.
\newblock Audiogen: Textually guided audio generation.
\newblock \emph{arXiv preprint arXiv:2209.15352}.

\bibitem[{Kubichek(1993)}]{407206}
R.~Kubichek. 1993.
\newblock \href {https://doi.org/10.1109/PACRIM.1993.407206} {Mel-cepstral distance measure for objective speech quality assessment}.
\newblock In \emph{Proceedings of IEEE Pacific Rim Conference on Communications Computers and Signal Processing}, volume~1, pages 125--128 vol.1.

\bibitem[{Kumar et~al.(2019)Kumar, Kumar, De~Boissiere, Gestin, Teoh, Sotelo, De~Brebisson, Bengio, and Courville}]{kumar2019melgan}
Kundan Kumar, Rithesh Kumar, Thibault De~Boissiere, Lucas Gestin, Wei~Zhen Teoh, Jose Sotelo, Alexandre De~Brebisson, Yoshua Bengio, and Aaron~C Courville. 2019.
\newblock Melgan: Generative adversarial networks for conditional waveform synthesis.
\newblock \emph{Advances in neural information processing systems}, 32.

\bibitem[{Kumar et~al.(2023)Kumar, Seetharaman, Luebs, Kumar, and Kumar}]{kumar2023high}
Rithesh Kumar, Prem Seetharaman, Alejandro Luebs, Ishaan Kumar, and Kundan Kumar. 2023.
\newblock High-fidelity audio compression with improved rvqgan.
\newblock \emph{arXiv preprint arXiv:2306.06546}.

\bibitem[{Lai et~al.(2021)Lai, Chuang, Lee, Li, and Glass}]{DBLP:conf/icassp/LaiCL0G21}
Cheng{-}I Lai, Yung{-}Sung Chuang, Hung{-}Yi Lee, Shang{-}Wen Li, and James~R. Glass. 2021.
\newblock Semi-supervised spoken language understanding via self-supervised speech and language model pretraining.
\newblock In \emph{{ICASSP}}, pages 7468--7472. {IEEE}.

\bibitem[{Lan et~al.(2023)Lan, Nagaraja, Chang, Kant, Ni, Shi, Iandola, and Chandra}]{lan2023stack}
Gael~Le Lan, Varun Nagaraja, Ernie Chang, David Kant, Zhaoheng Ni, Yangyang Shi, Forrest Iandola, and Vikas Chandra. 2023.
\newblock Stack-and-delay: a new codebook pattern for music generation.
\newblock \emph{arXiv preprint arXiv:2309.08804}.

\bibitem[{Lech et~al.(2020)Lech, Stolar, Best, and Bolia}]{lech2020real}
Margaret Lech, Melissa Stolar, Christopher Best, and Robert Bolia. 2020.
\newblock Real-time speech emotion recognition using a pre-trained image classification network: Effects of bandwidth reduction and companding.
\newblock \emph{Frontiers in Computer Science}, 2:14.

\bibitem[{Lugosch et~al.(2019)Lugosch, Ravanelli, Ignoto, Tomar, and Bengio}]{fluent}
Loren Lugosch, Mirco Ravanelli, Patrick Ignoto, Vikrant~Singh Tomar, and Yoshua Bengio. 2019.
\newblock Speech model pre-training for end-to-end spoken language understanding.
\newblock In \emph{Proc. of Interspeech}.

\bibitem[{Nagrani et~al.(2017)Nagrani, Chung, and Zisserman}]{DBLP:conf/interspeech/NagraniCZ17}
Arsha Nagrani, Joon~Son Chung, and Andrew Zisserman. 2017.
\newblock Voxceleb: {A} large-scale speaker identification dataset.
\newblock In \emph{{INTERSPEECH}}, pages 2616--2620. {ISCA}.

\bibitem[{Panayotov et~al.(2015)Panayotov, Chen, Povey, and Khudanpur}]{DBLP:conf/icassp/PanayotovCPK15}
Vassil Panayotov, Guoguo Chen, Daniel Povey, and Sanjeev Khudanpur. 2015.
\newblock Librispeech: An {ASR} corpus based on public domain audio books.
\newblock In \emph{{ICASSP}}, pages 5206--5210. {IEEE}.

\bibitem[{Piczak(2015)}]{piczak2015dataset}
Karol~J. Piczak. 2015.
\newblock \href {https://doi.org/10.1145/2733373.2806390} {{ESC}: {Dataset} for {Environmental Sound Classification}}.
\newblock In \emph{Proceedings of the 23rd {Annual ACM Conference} on {Multimedia}}, pages 1015--1018. {ACM Press}.

\bibitem[{Radford et~al.(2023)Radford, Kim, Xu, Brockman, McLeavey, and Sutskever}]{radford2023robust}
Alec Radford, Jong~Wook Kim, Tao Xu, Greg Brockman, Christine McLeavey, and Ilya Sutskever. 2023.
\newblock Robust speech recognition via large-scale weak supervision.
\newblock In \emph{International Conference on Machine Learning}, pages 28492--28518. PMLR.

\bibitem[{Rix et~al.(2001)Rix, Beerends, Hollier, and Hekstra}]{rix2001perceptual}
Antony~W Rix, John~G Beerends, Michael~P Hollier, and Andries~P Hekstra. 2001.
\newblock Perceptual evaluation of speech quality (pesq)-a new method for speech quality assessment of telephone networks and codecs.
\newblock In \emph{2001 IEEE international conference on acoustics, speech, and signal processing. Proceedings (Cat. No. 01CH37221)}, volume~2, pages 749--752. IEEE.

\bibitem[{Rubenstein et~al.(2023)Rubenstein, Asawaroengchai, Nguyen, Bapna, Borsos, Quitry, Chen, Badawy, Han, Kharitonov et~al.}]{rubenstein2023audiopalm}
Paul~K Rubenstein, Chulayuth Asawaroengchai, Duc~Dung Nguyen, Ankur Bapna, Zal{\'a}n Borsos, F{\'e}lix de~Chaumont Quitry, Peter Chen, Dalia~El Badawy, Wei Han, Eugene Kharitonov, et~al. 2023.
\newblock Audiopalm: A large language model that can speak and listen.
\newblock \emph{arXiv preprint arXiv:2306.12925}.

\bibitem[{Snyder et~al.(2018)Snyder, Garcia-Romero, Sell, Povey, and Khudanpur}]{snyder2018x}
David Snyder, Daniel Garcia-Romero, Gregory Sell, Daniel Povey, and Sanjeev Khudanpur. 2018.
\newblock X-vectors: Robust dnn embeddings for speaker recognition.
\newblock In \emph{2018 IEEE international conference on acoustics, speech and signal processing (ICASSP)}, pages 5329--5333. IEEE.

\bibitem[{St{\"o}ter et~al.(2018)St{\"o}ter, Chakrabarty, Edler, and Habets}]{stoter2018classification}
Fabian-Robert St{\"o}ter, Soumitro Chakrabarty, Bernd Edler, and Emanu{\"e}l~AP Habets. 2018.
\newblock Classification vs. regression in supervised learning for single channel speaker count estimation.
\newblock In \emph{2018 IEEE International Conference on Acoustics, Speech and Signal Processing (ICASSP)}, pages 436--440. IEEE.

\bibitem[{Taal et~al.(2010)Taal, Hendriks, Heusdens, and Jensen}]{taal2010short}
Cees~H Taal, Richard~C Hendriks, Richard Heusdens, and Jesper Jensen. 2010.
\newblock A short-time objective intelligibility measure for time-frequency weighted noisy speech.
\newblock In \emph{2010 IEEE international conference on acoustics, speech and signal processing}, pages 4214--4217. IEEE.

\bibitem[{Tagliasacchi et~al.(2020)Tagliasacchi, Li, Misiunas, and Roblek}]{tagliasacchi2020seanet}
Marco Tagliasacchi, Yunpeng Li, Karolis Misiunas, and Dominik Roblek. 2020.
\newblock Seanet: A multi-modal speech enhancement network.
\newblock \emph{arXiv preprint arXiv:2009.02095}.

\bibitem[{Tzanetakis and Cook(2002)}]{tzanetakis2002musical}
George Tzanetakis and Perry Cook. 2002.
\newblock Musical genre classification of audio signals.
\newblock \emph{IEEE Transactions on speech and audio processing}, 10(5):293--302.

\bibitem[{Valk and Alum{\"a}e(2021)}]{valk2021slt}
J{\"o}rgen Valk and Tanel Alum{\"a}e. 2021.
\newblock {VoxLingua107}: a dataset for spoken language recognition.
\newblock In \emph{Proc. IEEE SLT Workshop}.

\bibitem[{Vaswani et~al.(2017)Vaswani, Shazeer, Parmar, Uszkoreit, Jones, Gomez, Kaiser, and Polosukhin}]{vaswani2017attention}
Ashish Vaswani, Noam Shazeer, Niki Parmar, Jakob Uszkoreit, Llion Jones, Aidan~N Gomez, {\L}ukasz Kaiser, and Illia Polosukhin. 2017.
\newblock Attention is all you need.
\newblock \emph{Advances in neural information processing systems}, 30.

\bibitem[{Wang et~al.(2023{\natexlab{a}})Wang, Chen, Wu, Zhang, Zhou, Liu, Chen, Liu, Wang, Li et~al.}]{wang2023neural}
Chengyi Wang, Sanyuan Chen, Yu~Wu, Ziqiang Zhang, Long Zhou, Shujie Liu, Zhuo Chen, Yanqing Liu, Huaming Wang, Jinyu Li, et~al. 2023{\natexlab{a}}.
\newblock Neural codec language models are zero-shot text to speech synthesizers.
\newblock \emph{arXiv preprint arXiv:2301.02111}.

\bibitem[{Wang et~al.(2023{\natexlab{b}})Wang, Lee, Jang, and Su}]{wang2023zero}
Jun-You Wang, Hung-Yi Lee, Jyh-Shing~Roger Jang, and Li~Su. 2023{\natexlab{b}}.
\newblock Zero-shot singing voice synthesis from musical score.
\newblock In \emph{2023 IEEE Automatic Speech Recognition and Understanding Workshop (ASRU)}, pages 1--8. IEEE.

\bibitem[{Wang et~al.(2023{\natexlab{c}})Wang, Zhou, Zhang, Wu, Liu, Gaur, Chen, Li, and Wei}]{wang2023viola}
Tianrui Wang, Long Zhou, Ziqiang Zhang, Yu~Wu, Shujie Liu, Yashesh Gaur, Zhuo Chen, Jinyu Li, and Furu Wei. 2023{\natexlab{c}}.
\newblock Viola: Unified codec language models for speech recognition, synthesis, and translation.
\newblock \emph{arXiv preprint arXiv:2305.16107}.

\bibitem[{Wang et~al.(2023{\natexlab{d}})Wang, Thakker, Chen, Kanda, Eskimez, Chen, Tang, Liu, Li, and Yoshioka}]{wang2023speechx}
Xiaofei Wang, Manthan Thakker, Zhuo Chen, Naoyuki Kanda, Sefik~Emre Eskimez, Sanyuan Chen, Min Tang, Shujie Liu, Jinyu Li, and Takuya Yoshioka. 2023{\natexlab{d}}.
\newblock Speechx: Neural codec language model as a versatile speech transformer.
\newblock \emph{arXiv preprint arXiv:2308.06873}.

\bibitem[{Warden(2017)}]{speechcommands}
Pete Warden. 2017.
\newblock Speech commands: A public dataset for single-word speech recognition.
\newblock \emph{Dataset available online}.

\bibitem[{Wilkins et~al.(2018)Wilkins, Seetharaman, Wahl, and Pardo}]{Wilkins2018VocalSetAS}
Julia Wilkins, Prem Seetharaman, Alison Wahl, and Bryan Pardo. 2018.
\newblock \href {https://api.semanticscholar.org/CorpusID:53875542} {Vocalset: A singing voice dataset}.
\newblock In \emph{International Society for Music Information Retrieval Conference}.

\bibitem[{Wu et~al.(2023)Wu, Gebru, Markovi{\'c}, and Richard}]{wu2023audiodec}
Yi-Chiao Wu, Israel~D Gebru, Dejan Markovi{\'c}, and Alexander Richard. 2023.
\newblock Audiodec: An open-source streaming high-fidelity neural audio codec.
\newblock In \emph{ICASSP 2023-2023 IEEE International Conference on Acoustics, Speech and Signal Processing (ICASSP)}, pages 1--5. IEEE.

\bibitem[{Yang et~al.(2023{\natexlab{a}})Yang, Liu, Huang, Tian, Weng, and Zou}]{yang2023hifi}
Dongchao Yang, Songxiang Liu, Rongjie Huang, Jinchuan Tian, Chao Weng, and Yuexian Zou. 2023{\natexlab{a}}.
\newblock Hifi-codec: Group-residual vector quantization for high fidelity audio codec.
\newblock \emph{arXiv preprint arXiv:2305.02765}.

\bibitem[{Yang et~al.(2023{\natexlab{b}})Yang, Tian, Tan, Huang, Liu, Chang, Shi, Zhao, Bian, Wu et~al.}]{yang2023uniaudio}
Dongchao Yang, Jinchuan Tian, Xu~Tan, Rongjie Huang, Songxiang Liu, Xuankai Chang, Jiatong Shi, Sheng Zhao, Jiang Bian, Xixin Wu, et~al. 2023{\natexlab{b}}.
\newblock Uniaudio: An audio foundation model toward universal audio generation.
\newblock \emph{arXiv preprint arXiv:2310.00704}.

\bibitem[{Zeghidour et~al.(2021)Zeghidour, Luebs, Omran, Skoglund, and Tagliasacchi}]{zeghidour2021soundstream}
Neil Zeghidour, Alejandro Luebs, Ahmed Omran, Jan Skoglund, and Marco Tagliasacchi. 2021.
\newblock Soundstream: An end-to-end neural audio codec.
\newblock \emph{IEEE/ACM Transactions on Audio, Speech, and Language Processing}, 30:495--507.

\bibitem[{Zen et~al.(2019)Zen, Dang, Clark, Zhang, Weiss, Jia, Chen, and Wu}]{zen2019libritts}
Heiga Zen, Viet Dang, Rob Clark, Yu~Zhang, Ron~J Weiss, Ye~Jia, Zhifeng Chen, and Yonghui Wu. 2019.
\newblock Libritts: A corpus derived from librispeech for text-to-speech.
\newblock \emph{arXiv preprint arXiv:1904.02882}.

\bibitem[{Zhang et~al.(2022)Zhang, Li, Wang, Deng, Liu, Ren, He, Huang, Zhu, Chen, and Zhao}]{zhang2022msinger}
Lichao Zhang, Ruiqi Li, Shoutong Wang, Liqun Deng, Jinglin Liu, Yi~Ren, Jinzheng He, Rongjie Huang, Jieming Zhu, Xiao Chen, and Zhou Zhao. 2022.
\newblock M4singer: A multi-style, multi-singer and musical score provided mandarin singing corpus.
\newblock In \emph{Thirty-sixth Conference on Neural Information Processing Systems Datasets and Benchmarks Track}.

\bibitem[{Zhang et~al.(2023{\natexlab{a}})Zhang, Zhang, Li, Zhou, and Qiu}]{zhang2023speechtokenizer}
Xin Zhang, Dong Zhang, Shimin Li, Yaqian Zhou, and Xipeng Qiu. 2023{\natexlab{a}}.
\newblock Speechtokenizer: Unified speech tokenizer for speech large language models.
\newblock \emph{arXiv preprint arXiv:2308.16692}.

\bibitem[{Zhang et~al.(2023{\natexlab{b}})Zhang, Zhou, Wang, Chen, Wu, Liu, Chen, Liu, Wang, Li et~al.}]{zhang2023speak}
Ziqiang Zhang, Long Zhou, Chengyi Wang, Sanyuan Chen, Yu~Wu, Shujie Liu, Zhuo Chen, Yanqing Liu, Huaming Wang, Jinyu Li, et~al. 2023{\natexlab{b}}.
\newblock Speak foreign languages with your own voice: Cross-lingual neural codec language modeling.
\newblock \emph{arXiv preprint arXiv:2303.03926}.

\bibitem[{Ziyin et~al.(2020)Ziyin, Hartwig, and Ueda}]{ziyin2020neural}
Liu Ziyin, Tilman Hartwig, and Masahito Ueda. 2020.
\newblock Neural networks fail to learn periodic functions and how to fix it.
\newblock \emph{Advances in Neural Information Processing Systems}, 33:1583--1594.

\end{thebibliography}

\newpage
\appendix
\section{Technical Appendix}
\subsection{Dataset description}
\label{subsec: Dataset description}
21 public datasets are adopted in this work for zero-shot evaluation, including 9 datasets for speech, 4 datasets for audio, and 8 datasets for music. If not specified, we use the whole dataset for evaluation. All the dataset licenses are shown in Table~\ref{tab:dataset_license}.

\subsubsection{Speech}
\noindent\textbf{Speech Commands v1} Google Speech Command v1 \citep{speechcommands} is a dataset designed for recognizing spoken commands, consisting of 64,727 utterances from 1,881 speakers, with each utterance normalized as a 1-second waveform.

\noindent\textbf{QUESST} The QUESST 2014 dataset \citep{anguera2015quesst2014} contains 23 hours of spoken documents in six low-resource languages, encoded at 8 KHz and 16-bit resolution, sourced from various speech types and acoustic environments. 

\noindent\textbf{Fluent Speech Commands} The Fluent Speech Commands dataset \citep{fluent} comprises 30,043 spoken utterances from 97 individuals, recorded as single-channel .wav files at a 16 kHz sampling rate. Each file captures a distinct utterance intended for the operation of smart-home devices or a virtual assistant. For example, an utterance might be ``turn on the light in the bedroom.'' We use the test set for codec evaluation.

\noindent\textbf{LibriSpeech} LibriSpeech \citep{DBLP:conf/icassp/PanayotovCPK15} is a highly utilized corpus of English speech data, comprising roughly 1000 hours of audio recordings. These recordings are characterized by a reading style, as they consist of utterances read from audiobooks. We use test-clean and test-other sets for codec evaluation.

\noindent\textbf{Audio SNIPS} The Audio SNIPS dataset \citep{DBLP:conf/icassp/LaiCL0G21} utilizes a text-to-speech (TTS) system to synthesize the SNIPS dataset into utterances with different speakers and accents. The dataset is designed for speech recognition and natural language understanding simultaneously. We use test and valid splits for codec evaluation.

\noindent\textbf{VoxCeleb1} VoxCeleb \citep{DBLP:conf/interspeech/NagraniCZ17} is an audio-visual dataset featuring short segments of human speech sourced from interview videos on YouTube. It includes over a million real-world utterances from more than 6000 speakers. We use the test set for codec evaluation.

\noindent\textbf{IEMOCAP} The IEMOCAP dataset \citep{busso2008iemocap}, aimed at Multimodal Emotion Recognition, comprises 151 dialogue recordings, amounting to 302 videos due to the presence of two speakers in each session. It features annotations for 9 distinct emotions (angry, excited, fear, sad, surprised, frustrated, happy, disappointed, and neutral) and valence, arousal, and dominance. 

\noindent\textbf{Libri2Mix} Libri2Mix \citep{cosentino2020librimix} is a synthesized corpus featuring mixtures of two speakers' speech intertwined with background noise. The speech segments are sourced from LibriSpeech, while the ambient noise is taken from the WHAM! dataset. The corpus is organized into four subsets: train-360, train-100, dev, and test, cumulatively encompassing 300 hours of speech. We use the test set for codec evaluation.

\noindent\textbf{CREMA-D} \citep{6849440} is a 7,442 original clips from 91 actors (48 male and 43 female). Each clip is annotated with six distinct emotions. 
The professional actors, guided by experienced theatre directors, skillfully express a designated emotion while delivering specific sentences.

\noindent\textbf{LibriCount} \citep{stoter2018classification} is a generated dataset where each audio clip simulates a cocktail party scenario, incorporating 0 to 10 speech segments from LibriSpeech Test-Clean mixed with a signal-to-noise ratio (SNR) of 0dB.

\noindent\textbf{VoxLingua107 Top 10} \citep{valk2021slt} comprises audio segments for spoken language identification, encompassing 107 distinct languages. 
The audio clips in this dataset are automatically extracted from YouTube videos. 
We use the audio clips from the top 10 most frequent languages. 

\subsubsection{Audio}

\noindent\textbf{ESC-50} \citep{piczak2015dataset} encompasses 2000 environmental sounds categorized into 50 classes. 
The clips within this dataset are manually selected from public field recordings compiled by the Freesound.org project.

\noindent\textbf{FSD50K} \citep{fonseca2022FSD50K} is an open collection of human-labeled sound events. It comprises 51,197 Freesound clips distributed across 200 classes, selected from the AudioSet Ontology. We use a test and valid set for codec evaluation.

\noindent\textbf{Gunshot Triangulation} \citep{cooper2020gunshots}  collect the audio of seven distinct firearms—comprising four pistols and three rifles—each fired a minimum of three times. The shots were directed sequentially toward a target positioned 45 meters away from the shooter in an open field. The sound associated with these firings was captured using four separate iPod Touch devices.

\noindent\textbf{Vocal Imitations} \citep{kim2018vocal} comprises 11,242 crowd-sourced vocal imitations covering 302 sound event classes. The original sound recordings for these classes were sourced from Freesound, while their corresponding imitations were gathered through crowd-sourcing. We use the human imitation samples in the "included" split. 

\subsubsection{Music}
\noindent\textbf{OpenSinger} \citep{huang2021multi}, a Chinese multi-singer vocal dataset, features high-fidelity recordings by professional singers, free of noise and background interference. OpenSinger does not have a standard way of splitting datasets. 
We use the songs with male prefixes from 25 to 27 and female with prefixes from 45 to 47 in the dataset as our test set, which is based on the split method of a recent paper on zero-short singing voice synthesis \citep{wang2023zero}. We use the test set for codec evaluation.

\noindent\textbf{M4Singer} \citep{zhang2022msinger} offers a rich collection of about 700 Chinese pop songs recorded by 20 professional vocalists, encompassing all four SATB voice types: soprano, alto, tenor, and bass. Following UniAudio \citep{yang2023uniaudio}, we conduct experiments on the M4Singer test set. We use the test set for codec evaluation.  

\noindent\textbf{VocalSet} \citep{Wilkins2018VocalSetAS} comprises 10.1 hours of recordings from 20 professional singers (11 male, 9 female), executing 17 distinct vocal techniques, which aids in the development of advanced machine learning models for applications like singer identification, vocal technique detection, and singing synthesis. We only use the test set of VocalSet for experiments.

\noindent\textbf{NSynth} \citep{pmlr-v70-engel17a} stands out as a large-scale, high-quality collection of musical notes, significantly surpassing similar public datasets in size. 
We only use the test set of the NSynth dataset for experiments.


\noindent\textbf{GTZAN Genre} \citep{tzanetakis2002musical} includes music samples categorized into 10 genres, each containing 100 audio files. All audio files within the dataset have a standardized length of 30 seconds. 

\noindent\textbf{GTZAN Music Speech} \citep{tzanetakis2002musical} consists of both music and speech segments, with each category containing 60 samples having a standardized length of 30 seconds.


\begin{table*}[h]
\centering

\begin{tabular}{@{}llllll@{}}
\toprule
\textbf{Speech Dataset}   & \textbf{License} \\
\midrule
LibriSpeech & CC BY 4.0 \\
VoxCeleb1 & CC BY 4.0 \\
Speech Commands v1 & CC BY 4.0  \\ 
QUESST &  Free For Research Purposes \\
VoxLingua107 Top 10 & CC BY 4.0  \\
Audio SNIPS & CC0 v1.0 \\
IEMOCAP & SAIL Agreement \\
CREMA-D &  Open Database License \\
Libri2Mix & MIT License \\
LibriCount &  CC BY 4.0 \\
\midrule
\textbf{Audio Dataset}   & \textbf{License} \\
\midrule
ESC-50 & CC BY-NC 3.0  \\
FSD-50K & CC0, CC BY 4.0, CC BY-NC 4.0, CC Sampling+ 1.0  \\
Gunshot Triangulation & CC0 \\
Vocal Imitations & CC BY 4.0 \\
\midrule
\textbf{Music Dataset}   & \textbf{License} \\
\midrule
OpenSinger & CC BY-NC-SA 2.0 \\ 
M4Singer & CC BY-NC-SA 4.0 \\ 
VocalSet & CC BY 4.0 \\ 
NSynth & CC BY 4.0 \\
GTZAN Genre & CC BY 4.0, Apache License v.2.0 \\
GTZAN Music Speech & CC BY 4.0, Apache License v.2.0 \\
\bottomrule
\end{tabular}
\caption{The dataset license statistics. }
\label{tab:dataset_license}
\end{table*}

\subsection{Downstream task description}
\label{sub_appendix: Downstream task description}

\subsubsection{Automatic Speech Recognition (ASR)}
\label{subsub_appendix: ASR}
ASR is an essential component in speech processing, aiming to convert speech into text. 
ASR is dedicated to extracting and interpreting the content information embedded in speech. 
This involves understanding various linguistic elements within the spoken language, such as phonetics, syntax, and semantics. 
ASR systems are instrumental in numerous applications, including voice-activated assistants, transcription services, and interactive voice response systems, where accurate content interpretation is crucial.

Our study employs the Whisper model \citep{radford2023robust}, specifically the ``whisper-large" variant, the current state-of-the-art ASR system, for evaluation. 
Whisper stands out due to its robust and versatile architecture, which is capable of handling a broad range of speech recognition tasks across multiple languages. 
The model's foundation is an encoder-decoder Transformer, adept at learning from large, weakly supervised datasets. 
This enables Whisper to perform effectively in diverse scenarios without requiring specific dataset fine-tuning. 
The model's comprehensive training approach, focusing on generalization and robustness, positions it as a powerful tool for speech content interpretation.

To evaluate the performance of the Whisper model in ASR, we utilize the metric Word Error Rate (WER). 
WER measures the percentage of errors at the word level, offering insights into the model's accuracy in transcribing speech to text. 
The metric is critical in assessing the effectiveness of ASR systems, allowing for a detailed understanding of their capabilities in accurately capturing and converting spoken language into written form.
In this evaluation, we analyze the ASR performance using subsets of the LibriSpeech dataset, precisely the test-clean and test-other subsets, to ensure a comprehensive assessment of the model's transcription accuracy.

\subsubsection{Automatic Speaker Verification (ASV)}
\label{subsub_appendix: ASV}
In contrast to text, which primarily conveys content information, speaker information represents a distinct and unique aspect of speech. 
We employ ASV to assess the degree of speaker information loss in the resynthesized speech generated by neural codecs.
ASV is a cutting-edge technology that plays a pivotal role in voice authentication and security systems. 
ASV is designed to verify a person's claimed identity by analyzing their unique vocal characteristics, such as pitch, tone, and speech patterns. 
It offers a seamless and secure method of confirming whether an individual is who they claim to be, making it a valuable tool in applications ranging from access control and secure transactions to law enforcement and customer service.

We utilize the cutting-edge speaker verification model, ECAPA-TDNN \citep{desplanques2020ecapa}, which is pre-trained on the VoxCeleb2 dataset \citep{chung2018voxceleb2}, as the pre-trained ASV model.
Building upon the well-established x-vector architecture \citep{snyder2018x}, ECAPA-TDNN introduces several novel enhancements inspired by recent trends in face verification.

We adopt equal error rate (EER) as two evaluation metrics to evaluate the performance of ASV. 
EER provides a balance between false acceptances and rejections, and minDCF allows for a more nuanced assessment of system performance by considering the costs associated with different types of errors (false acceptances and rejections).

\subsubsection{Emotion recognition (ER)}
\label{subsub_appendix: ER}
In addition to speaker information, speech conveys affective information, including emotions. 
We employ ER to quantify the degree of paralinguistic information loss due to speech resynthesis by codec models.
ER is an essential component in human-computer interaction, such as smart entertainment, healthcare, or e-learning.
ER specifically identifies the emotional components of speech that are unrelated to semantic information \citep{lech2020real}.

We adopt the state-of-the-art self-supervised model WavLM-Large \citep{chen2022wavlm} as feature extractor, and train the downstream emotion classification model using the weighted sum of hidden states as the representation. 
Following the SUPERB benchmark, we employ mean-pooling followed by a linear layer for modeling purposes and utilize cross-entropy as the training loss function. 
As for the dataset, we select a subset of the IEMOCAP dataset \citep{busso2008iemocap}, where we have excluded the unbalanced emotion classes, resulting in four remaining classes (neutral, happy, sad, angry). 
We further divided this subset into five folds for cross-validation purposes.
We report the average classification accuracy across the five folds.

\subsubsection{Audio event classification}
\label{subsub_appendix: AEC}

ASE aims to automatically identify and categorize specific sound events or occurrences within an audio recording. 
These sound events can be various sounds, such as footsteps, car horns, dog barks, music genres, or any other acoustic events. 
We use AudioSet \citep{45857} as the evaluation set.
AudioSet offers a comprehensive library of sound events, categorized in a hierarchical structure that spans a broad spectrum of sounds, from human and animal noises to natural and environmental sounds and musical and miscellaneous audio events.
The Audio Spectrogram Transformer (AST), proposed by \cite{gong21b_interspeech}, is a high-performance open-source AEC model. 
The model takes spectrograms as input features, divides them into patch embeddings, and adds a learnable position embedding for each patch.
An extra classification token is added to the input sequence at the beginning. 
Subsequently, the feature sequence is fed into a Transformer Encoder to make predictions.

We employ the pre-trained AST model\footnote{\href{https://github.com/YuanGongND/ast}{https://github.com/YuanGongND/ast}} for our AEC downstream task evaluation.
We adopt mean average precision (mAP)\footnote{\href{https://scikit-learn.org/stable/modules/generated/sklearn.metrics.average_precision_score.html}{https://scikit-learn.org/}} to evaluate the AEC performance.
The AST model had undergone pre-training on the AudioSet training dataset and had been subjected to post-processing using a weight averaging strategy to obtain 45.9 mAP(\%) at its evaluation set. 

\subsection{Signal-level metrics}
\label{sub_appendix: Objective metrics}
Aligned with expertise in the speech domain, we assess codec models using a comprehensive set of Signal-level metrics, encompassing various aspects of audio quality. These include:
\begin{itemize}
    \item \textbf{STFTDistance}: Evaluates frequency content by calculating the L1-loss over multi-scale STFT (Short-Time Fourier Transform) representations, measuring frequency discrepancies across multiple resolutions. This method provides a detailed assessment of frequency content and temporal dynamics.
    \item \textbf{MelDistance}: Employs the L1-loss between log Mel spectrogram representations to gauge the fidelity of spectral features, reflecting spectral fidelity and timbral texture in the audio.
    \item \textbf{PESQ}: Perceptual Evaluation of Speech Quality, Rates the perceptual quality of speech, mimicking human auditory perception to provide a subjective quality score.
    \item \textbf{STOI}: Short-Time Objective Intelligibility, Evaluates speech intelligibility, especially in noisy conditions, ensuring clarity and comprehensibility of the generated speech
    \item \textbf{F0CORR (F0 Pearson Correlation Coefficient)}: Evaluates the pitch accuracy between original and synthesized audio by aligning their fundamental frequency (F0) contours using dynamic time warping (DTW) and then computing the Pearson correlation. This metric highlights the codec's ability to maintain pitch, which is essential for audio naturalness and expressiveness.
\end{itemize}
These metrics provide a comprehensive evaluation of codec models, focusing on both quantitative accuracy and perceptual quality of audio.

\subsection{Codec models}
\label{sub_appendix: Codec models}
\noindent\textbf{SoundStream:} SoundStream \citep{zeghidour2021soundstream} stands as one of the pioneering implementations of neural codec models, embodying a classic neural codec architecture comprising encoder, quantizer, and decoder modules.
It utilizes the streaming SEANets \citep{tagliasacchi2020seanet} as its encoder and decoder. 
The quantizer incorporates a speech enhancement system with a Residual Vector Quantization (RVQ) \citep{kumar2019melgan, zeghidour2021soundstream} bottleneck to obtain parallel token streams. 
During training, the model parameters are optimized using a combination of reconstruction and adversarial loss. 
SoundStorm \citep{borsos2023soundstorm} is an improved version of SoundStream to achieve both efficiency and high-quality audio generation. 
It accomplishes this by employing an architecture specifically tailored to the hierarchical structure of audio tokens. 
Moreover, it pioneers a parallel, non-autoregressive decoding scheme, which relies on confidence-based strategies for residual vector-quantized token sequences. 

\noindent\textbf{Encodec:} Encodec \citep{defossez2022high} builds upon a framework similar to SoundStream. 
Nonetheless, it further augments its capabilities by integrating supplementary LSTM \citep{hochreiter1997long} layers and harnessing a Transformer-based language model \citep{vaswani2017attention} to model the RVQ codes, thereby amplifying its sequence modeling performance.
Then, there is a stream of work aimed at making codec models more general and powerful.
AudioDec \citep{wu2023audiodec} represents an enhanced version of Encodec, implementing a group convolution mechanism to facilitate real-time operation of the streamable network, while also harnessing the capabilities of HiFi-GAN \citep{kong2020hifi} to effectively generate high-fidelity audio at a sampling rate of 48 kHz.

\noindent\textbf{AcademiCodec:} In the AcademiCodec model introduced by \cite{yang2023hifi}, a novel technique known as group-residual vector quantization is presented. 
This technique is tailored explicitly for generation tasks. It aims to enhance the reconstruction performance using a limited number of codebooks, consequently achieving an impressively low bit rate per second (BPS). 
This low BPS is of utmost significance as it effectively addresses the challenge of lengthy speech tokens in speech-language modeling, resulting in reduced sequence lengths.

\noindent\textbf{SpeechTokenizer:} SpeechTokenizer \citep{zhang2023speechtokenizer} is a unified speech tokenizer designed for speech-language models. 
It implements an Encoder-Decoder architecture enhanced with RVQ. 
By integrating semantic and acoustic tokens, SpeechTokenizer hierarchically separates various facets of speech information across different RVQ layers.
Specifically, SpeechTokenizer is designed to regularize the first RVQ layer to learn the Hubert tokens \citep{hsu2021hubert}.
The authors claim that employing such techniques can lead to improved disentanglement of information across various RVQ layers.

\noindent\textbf{Descript-Audio-Codec:}
Descript-audio-codec (DAC) \citep{kumar2023high}, another instance of a universal neural codec model, distinguishes itself through its exceptional ability to maintain high-fidelity audio quality across a broad spectrum of data types, encompassing audio, music, and speech.
It accomplishes this feat by employing a multitude of training techniques, such as periodic activation functions \citep{ziyin2020neural}, enhanced residual vector quantization using factorized and L2-normalized codes, random quantizer dropout to preserve audio reconstruction quality, as well as refining adversarial and reconstruction loss during the training process.
Out of all the techniques employed, they emphasize the pivotal role played by the periodic activation function.

\noindent\textbf{FunCodec:} Unlike most models focusing on the time domain, FunCodec \citep{du2023funcodec} proposes a frequency-domain codec.
The authors claim they can achieve comparable performance with fewer parameters and lower computation complexity. 
Meanwhile, it also finds that incorporating semantic information in the codec tokens improves speech quality at low bit rates.

\section{Additional experiment results}
\label{appendix: Additional experiment results}
(Due to the space limitation, please refer to the next page.)
\begin{table*}[htbp]
\centering
\resizebox{1\linewidth}{!}{
\begin{tabular}{lccccccccccc}
\toprule
\textbf{Codec} & \textbf{Librispeech} & \textbf{Speech Commands} & \textbf{QUESST} & \textbf{VoxCeleb1} & \textbf{SNIPS} & \textbf{IEMOCAP} & \textbf{Libri2Mix} & \textbf{CREMA D} & \textbf{LibriCount} & \textbf{VoxLingua107Top10} & \textbf{Overall}\\
\midrule
A & 0.713 & 0.677 & 0.637 & 0.676 & 0.697 & 0.629 & 0.637 & 0.586 & 0.566 & 0.651 & 0.644 \\
\midrule
B1 & 0.679 & 0.633 & 0.615 & 0.632 & 0.666 & 0.579 & 0.589 & 0.550 & 0.544 & 0.641 & 0.610 \\
B2 & 0.678 & 0.638 & 0.625 & 0.637 & 0.655 & 0.587 & 0.601 & 0.568 & 0.555 & 0.647 & 0.617 \\
B3 & 0.673 & 0.633 & 0.618 & 0.631 & 0.662 & 0.584 & 0.592 & 0.555 & 0.549 & 0.643 & 0.611 \\
\midrule
C & 0.638 & 0.634 & 0.599 & 0.612 & 0.639 & 0.589 & 0.582 & 0.544 & 0.529 & 0.610 & 0.596 \\
\midrule
D1 & 0.824 & 0.822 & 0.801 & 0.822 & 0.808 & 0.820 & 0.814 & 0.764 & 0.748 & 0.764 & 0.798 \\
D2 & 0.873 & 0.876 & 0.907 & 0.874 & 0.868 & 0.898 & 0.876 & 0.834 & 0.822 & 0.818 & 0.864 \\
D3 & 0.805 & 0.810 & 0.807 & 0.808 & 0.794 & 0.815 & 0.796 & 0.807 & 0.778 & 0.802 & 0.802 \\
\midrule
E1 & 0.592 & 0.599 & 0.590 & 0.589 & 0.596 & 0.565 & 0.570 & 0.561 & 0.542 & 0.595 & 0.579 \\
E2 & 0.652 & 0.648 & 0.654 & 0.652 & 0.656 & 0.621 & 0.629 & 0.611 & 0.594 & 0.650 & 0.636 \\
E3 & 0.708 & 0.697 & 0.716 & 0.718 & 0.716 & 0.678 & 0.700 & 0.672 & 0.661 & 0.705 & 0.697 \\
E4 & 0.751 & 0.738 & 0.770 & 0.776 & 0.764 & 0.734 & 0.762 & 0.726 & 0.721 & 0.747 & 0.748 \\
E5 & 0.773 & 0.761 & 0.798 & 0.802 & 0.787 & 0.767 & 0.793 & 0.755 & 0.750 & 0.768 & 0.775 \\
\midrule
F1 & 0.764 & 0.732 & 0.730 & 0.748 & 0.764 & 0.749 & 0.740 & 0.672 & 0.656 & 0.702 & 0.724 \\
F2 & 0.759 & 0.726 & 0.706 & 0.729 & 0.771 & 0.733 & 0.707 & 0.641 & 0.608 & 0.694 & 0.704 \\
F3 & 0.750 & 0.732 & 0.708 & 0.747 & 0.775 & 0.731 & 0.677 & 0.648 & 0.609 & 0.704 & 0.705 \\
F4 & 0.736 & 0.700 & 0.690 & 0.723 & 0.753 & 0.684 & 0.656 & 0.607 & 0.587 & 0.686 & 0.678 \\
F5 & 0.781 & 0.747 & 0.738 & 0.776 & 0.793 & 0.750 & 0.698 & 0.671 & 0.618 & 0.723 & 0.726 \\
F6 & 0.773 & 0.736 & 0.735 & 0.769 & 0.769 & 0.732 & 0.696 & 0.661 & 0.617 & 0.723 & 0.718 \\
\bottomrule
\end{tabular}
}
\caption{Signal level overall scores across speech datasets}
\label{tab:combined_overall_scores_speech}
\end{table*}




\begin{table*}[htbp]
\centering
\fontsize{9}{10}\selectfont
\begin{tabular}{lccccc}
\toprule
\textbf{Codec} & \textbf{ESC-50} & \textbf{FSD50K} & \textbf{Gunshot Triangulation} & \textbf{Vocal Imitations} & \textbf{Overall}\\
\midrule
A & 0.575 & 0.577 & 0.594 & 0.580 & 0.581 \\
\midrule
B1 & 0.566 & 0.562 & 0.594 & 0.573 & 0.574 \\
B2 & 0.567 & 0.563 & 0.595 & 0.573 & 0.574 \\
B3 & 0.585 & 0.576 & 0.620 & 0.588 & 0.592 \\
\midrule
C & 0.599 & 0.597 & 0.614 & 0.598 & 0.602 \\
\midrule
D1 & 0.583 & 0.585 & 0.606 & 0.589 & 0.591 \\
D2 & 0.634 & 0.628 & 0.652 & 0.630 & 0.636 \\
D3 & 0.699 & 0.705 & 0.707 & 0.699 & 0.702 \\
\midrule
E1 & 0.596 & 0.592 & 0.605 & 0.585 & 0.594 \\
E2 & 0.600 & 0.595 & 0.611 & 0.589 & 0.599 \\
E3 & 0.604 & 0.599 & 0.614 & 0.593 & 0.602 \\
E4 & 0.609 & 0.602 & 0.618 & 0.595 & 0.606 \\
E5 & 0.612 & 0.604 & 0.623 & 0.597 & 0.609 \\
\midrule
F1 & 0.574 & 0.579 & 0.595 & 0.582 & 0.582 \\
F2 & 0.576 & 0.580 & 0.593 & 0.583 & 0.583 \\
F3 & 0.576 & 0.578 & 0.594 & 0.578 & 0.581 \\
F4 & 0.575 & 0.577 & 0.585 & 0.577 & 0.578 \\
F5 & 0.576 & 0.578 & 0.596 & 0.581 & 0.583 \\
F6 & 0.577 & 0.579 & 0.594 & 0.580 & 0.583 \\
\bottomrule
\end{tabular}
\caption{Signal level overall scores across audio datasets}
\label{tab:combined_overall_scores}
\end{table*}

\begin{table*}[htbp]
\centering
\fontsize{9}{10}\selectfont
\setlength\tabcolsep{3pt}
\begin{tabular}{lcccccc}
\toprule
\textbf{Codec} & \textbf{OpenSinger} & \textbf{m4singer} & \textbf{VocalSet} & \textbf{GTZAN Genre} & \textbf{GTZAN Music Speech} & \textbf{Overall}\\
\midrule
A & 0.717 & 0.711 & 0.565 & 0.481 & 0.531 & 0.585 \\
\midrule
B1 & 0.714 & 0.730 & 0.653 & 0.492 & 0.527 & 0.601 \\
B2 & 0.709 & 0.719 & 0.650 & 0.564 & 0.576 & 0.630 \\
B3 & 0.704 & 0.710 & 0.649 & 0.489 & 0.548 & 0.604 \\
\midrule
C & 0.680 & 0.679 & 0.579 & 0.473 & 0.514 & 0.572 \\
\midrule
D1 & 0.840 & 0.840 & 0.731 & 0.726 & 0.685 & 0.749 \\
D2 & 0.878 & 0.866 & 0.783 & 0.793 & 0.785 & 0.815 \\
D3 & 0.806 & 0.798 & 0.789 & 0.706 & 0.752 & 0.770 \\
\midrule
E1 & 0.628 & 0.616 & 0.519 & 0.499 & 0.580 & 0.568 \\
E2 & 0.680 & 0.663 & 0.581 & 0.548 & 0.632 & 0.621 \\
E3 & 0.724 & 0.699 & 0.625 & 0.589 & 0.697 & 0.669 \\
E4 & 0.762 & 0.730 & 0.658 & 0.623 & 0.756 & 0.710 \\
E5 & 0.784 & 0.749 & 0.675 & 0.642 & 0.784 & 0.732 \\
\midrule
F1 & 0.773 & 0.765 & 0.672 & 0.557 & 0.627 & 0.667 \\
F2 & 0.796 & 0.796 & 0.705 & 0.560 & 0.585 & 0.668 \\
F3 & 0.786 & 0.772 & 0.672 & 0.533 & 0.581 & 0.649 \\
F4 & 0.775 & 0.766 & 0.658 & 0.512 & 0.555 & 0.632 \\
F5 & 0.820 & 0.806 & 0.703 & 0.537 & 0.582 & 0.665 \\
F6 & 0.789 & 0.797 & 0.686 & 0.549 & 0.609 & 0.667 \\
\bottomrule
\end{tabular}
\caption{Signal level overall scores across music datasets}
\label{tab:combined_overall_scores_music}
\end{table*}

\begin{figure*}[h]
    \centering
    \includegraphics[width=1\linewidth]{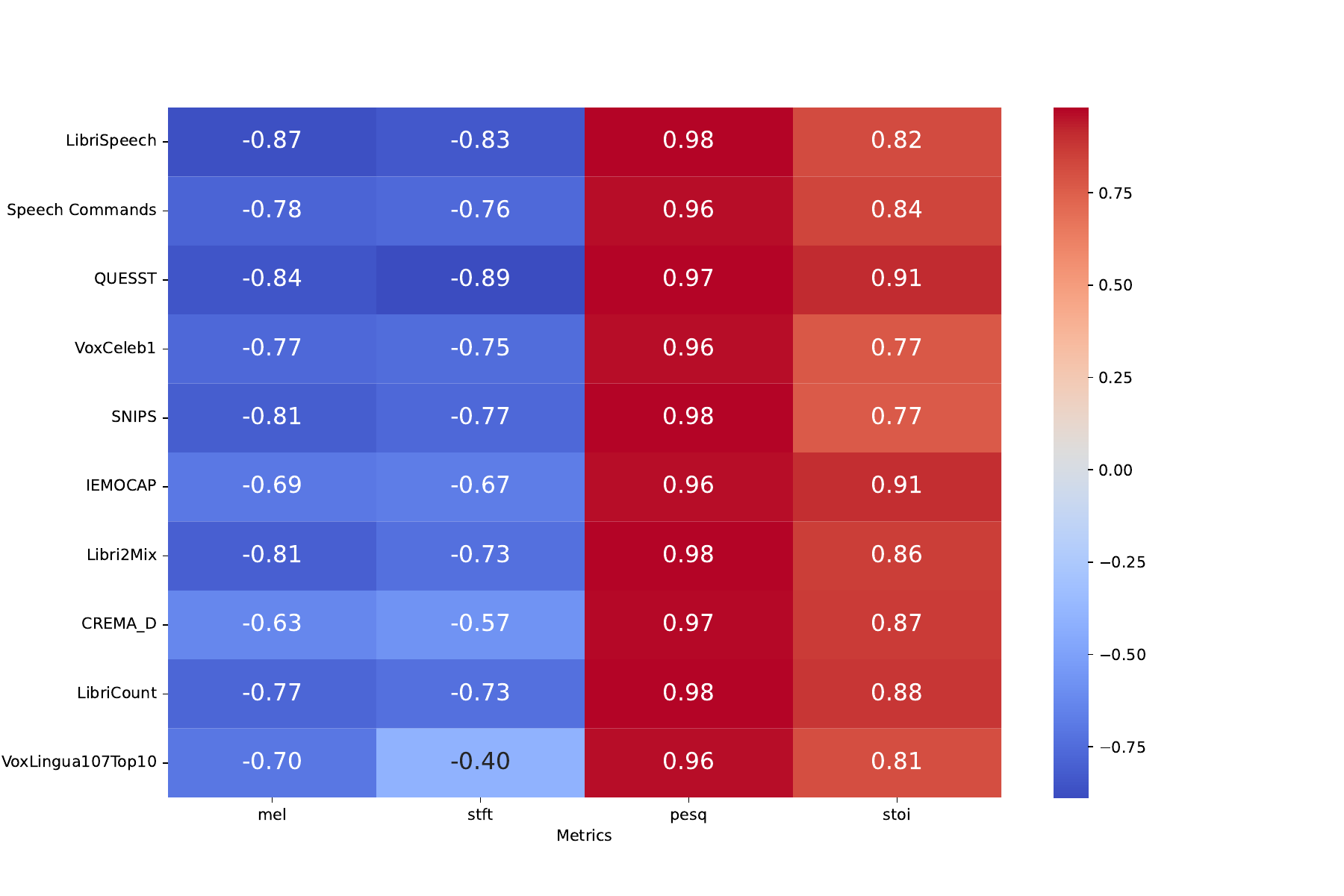}\hfill
    \caption{Speech overall score correlation}
    \label{fig:speech overall score correlation}
\end{figure*}

\begin{figure*}[h]
    \centering
    \includegraphics[width=1\linewidth]{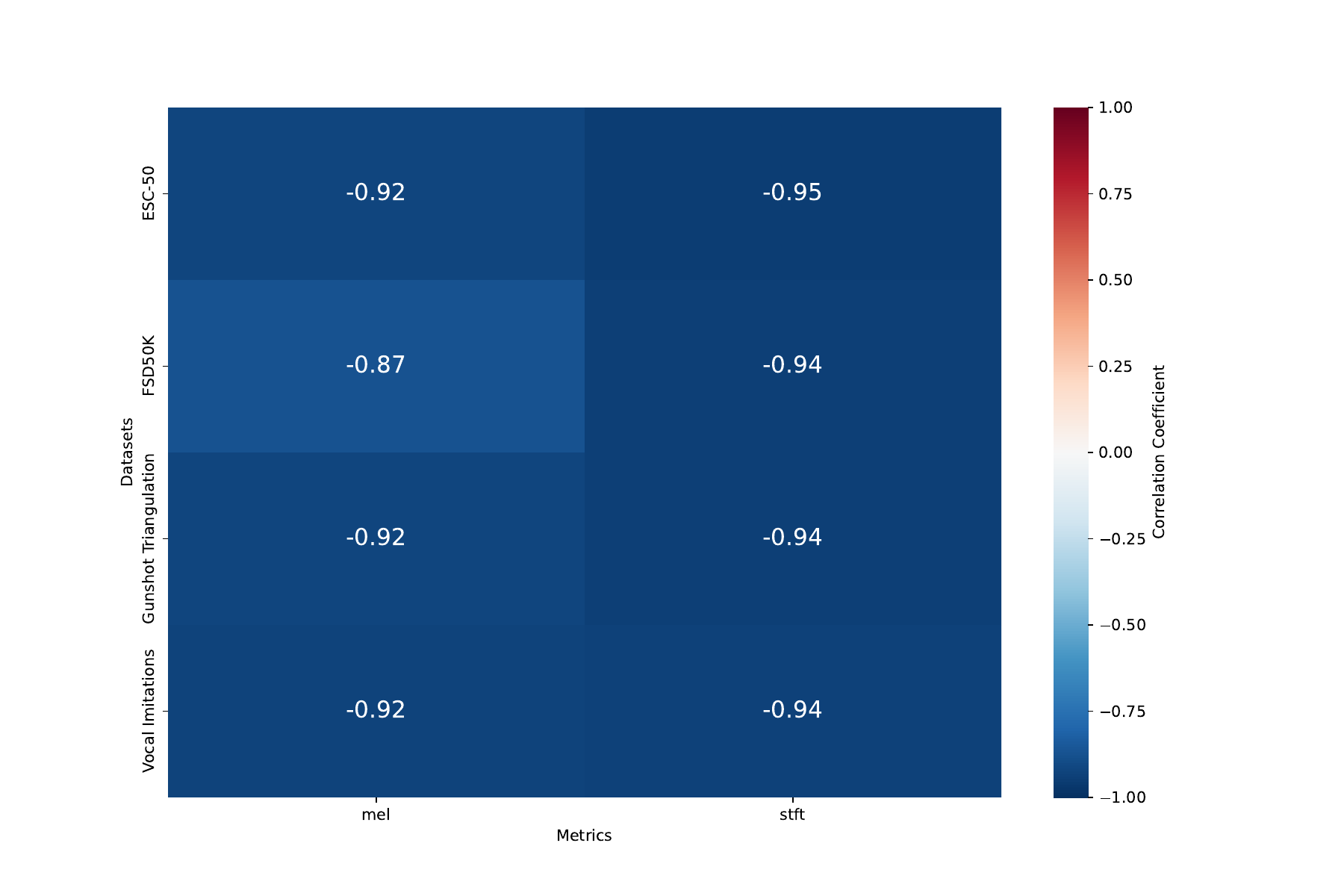}\hfill
    \caption{Audio overall score correlation}
    \label{fig:audio overall score correlation}
\end{figure*}



\begin{figure*}[h]
    \centering
    \includegraphics[width=1\linewidth]{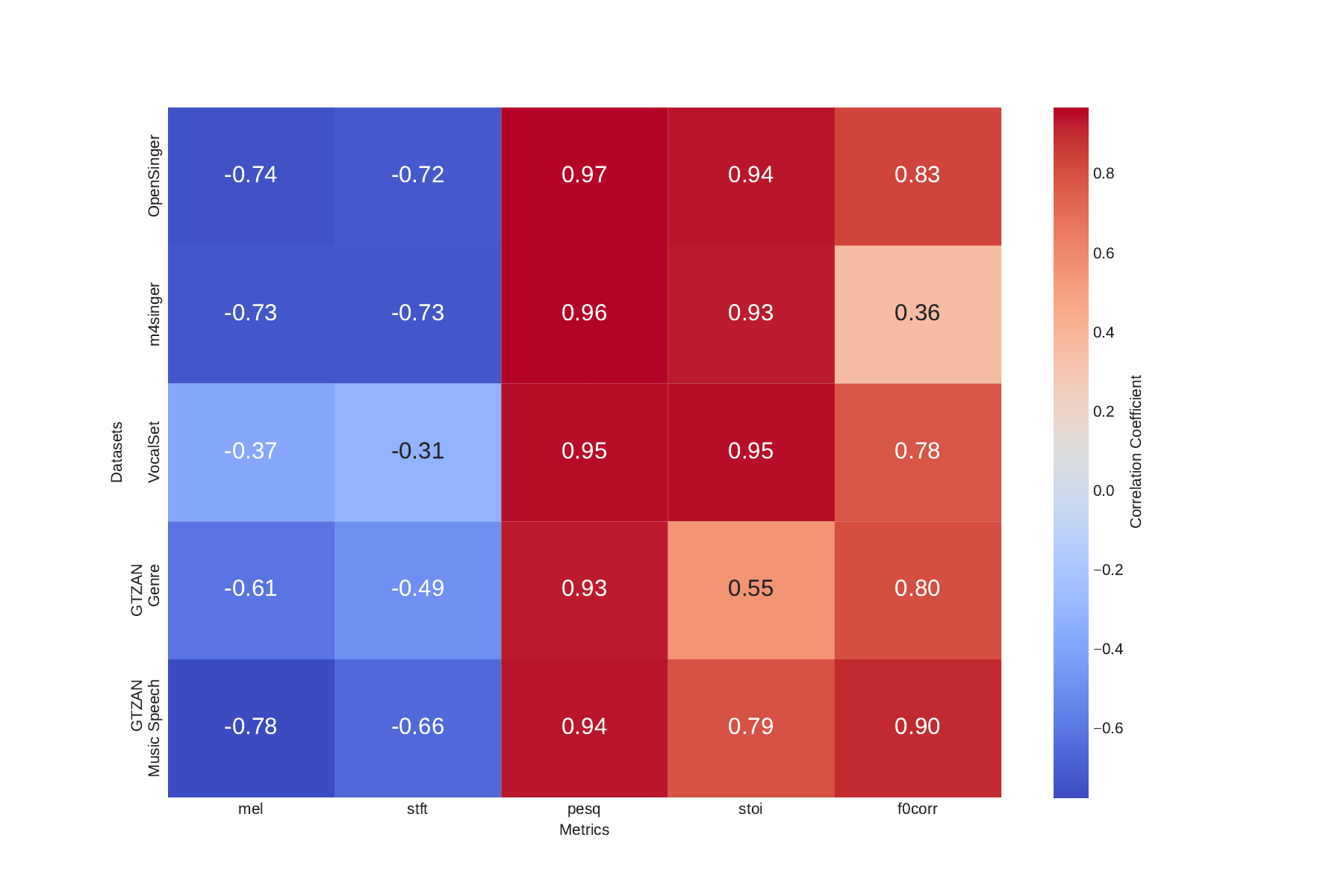}\hfill
    \caption{Music overall score correlation}
    \label{fig:music overall score correlation}
\end{figure*}

\end{document}